\documentclass[twocolumn,showpacs,preprintnumbers,amsmath,amssymb]{revtex4}
\usepackage{graphicx}
\usepackage{dcolumn}
\usepackage{bm}

\begin{document}

\title{Luttinger liquid theory of purple bronze $Li_{0.9}Mo_6 O_{17}$ in the charge regime.}

\author{P. Chudzinski}
\author{T. Jarlborg}
\author{T. Giamarchi}
\affiliation{
DPMC, University of Geneva, 24 Quai Ernest-Ansermet, CH-1211 Geneva 4,
Switzerland
}

\date{\today}

\begin{abstract}
Molybdenum purple bronze Li$_{0.9}$Mo$_{6}$O$_{17}$ is an
exceptional material known to exhibit one dimensional (1D)
properties for energies down to a few meV. This fact seems to be
well established both in experiments and in band structure theory.
We use the unusual, very 1-dimensional band dispersion obtained in
\emph{ab-initio} DFT-LMTO band calculations as our starting point
to study the physics emerging below 300meV. A dispersion
perpendicular to the main dispersive direction is obtained and
investigated in detail. Based on this, we derive an effective low
energy theory within the Tomonaga Luttinger liquid (TLL)
framework. We estimate the strength of the possible interactions
and from this deduce the values of the TLL parameters for charge
modes. Finally we investigate possible instabilities of TLL by
deriving renormalization group (RG) equations which allow us to
predict the size of potential gaps in the spectrum. While $2k_F$
instabilities strongly suppress each other, the $4k_F$
instabilities cooperate, which paves the way for a possible CDW at
the lowest energies. The aim of this work is to understand the
experimental findings, in particular the ones which are certainly
lying within the 1D regime. We discuss the validity of our 1D
approach and further perspectives for the lower energy phases.

\end{abstract}

\maketitle

\section{Introduction.}

The molybdenum purple bronze, $Li_{0.9}Mo_6 O_{17}$ , is a subject
of intensive experimental studies already for more than two
decades\cite{McCarroll-first}, but its unusual properties remain
unclear. Several very different experimental probes have been
used: angle resolved photo-emission (ARPES)\cite{JWAllen-old},
scanning tunnelling microscopy (STM)\cite{Hager-bron-STM},
DC\cite{Greenblatt-old-res-anis} and magneto-
resistivity\cite{Hussey-rhoB}, thermal conductivity
\cite{Hussey-thermo}, optical conductivity\cite{Mandrus-optics},
Nernst signal\cite{Santos-Nernst-2D}, muons
spectroscopy\cite{Chakhalian-muons},
X-rays\cite{Onoda-structure-Xray}, thermal
expansion\cite{Santos-thermal-expan}, neutron
scattering\cite{Santos-structure-neutrons}. Although the main
effort of those investigations was focused on the nature of a
mysterious phase transition at around 25K, interesting knowledge
about higher energy phase was also gathered. Certain properties of
the one dimensional (1D) metal, the Luttinger liquid (TLL), have
been invoked to explain the measured data\cite{JWAllen-old,
Hager-bron-STM, JWAllen-1D+2D, JWAllen-comment-Greenblatt} and
nowadays the presence of the 1D physics is well established
experimentally\cite{JWAllen-high-bulk, JWAllen-growth-meth}, at
least in some energy range.

On the theoretical side, band structure calculations have shown a
quasi-1D character of molybdenum purple bronze. A remarkably
simple band structure emerges from rather complex crystal
structure. At the Fermi surface there are only two bands, lying
very close to each other, in the form of flat sheets dispersing
well only along the b-axis. This gives a hope that purple bronze
can indeed be a rare realization of the 1D physics.

The key problem is that several possible mechanisms has been
invoked to explain the observed properties, which made the subject
quite unclear and controversial. In our opinion the reason for
this situation is that each of previous attempts was focused only
on one out of many peculiar properties of $Li_{0.9}Mo_6 O_{17}$
and most of them searched for an explanation in the low energy
regime (below $5meV$), where indeed the properties of
$Li_{0.9}Mo_6 O_{17}$ are the most spectacular. By now, not enough
attention has been paid even to the parameters of the 1D state. It
is only agreed that it emerges at energies as high as 250meV. The
values of these parameters are the first issue one must determine
before pursuing research towards low energy regimes. The unusual
physics observed in $Li_{0.9}Mo_6 O_{17}$ at the energy scales of
order 10meV is obviously a motivation for revising the question of
the 1D physics. In order to establish a proper low energy
effective theory one has to begin at highest energies and the step
by step move towards the physics taking place around Fermi energy.
The first step is to link the results of the DFT calculations with
the well defined field theory describing the experimental results
at around 20meV. It is this "high energy" regime which must be
well understood first. This is the main task of this paper.

The plan of this work is as follows. In Sec.\ref{sec:bands} we
begin with a brief introduction of the band structure. In the
following section Sec.\ref{ssec:tight-bind} we propose a tight
binding model which is able to approximate bands around the Fermi
energy $E_F$, however in addition it contains also the strong
correlation terms, beyond the LDA-DFT. In Sec.\ref{ssec:LL-intro}
we give basic notions of 1D physics, used in the rest of the
paper. Section Sec.\ref{sec:intra} is dedicated to intra-chain
physics: we give values parameterizing strong correlations
(Sec.\ref{sec:strint}), estimate TLL parameters which this implies
(Sec.\ref{sec:LL-res}) and give energy scales for the spin sector
(Sec.\ref{ssec:spin}). In section Sec.\ref{sec:inter-phys} we
introduce the inter-chain physics. Once again, first we estimate
the strength of these interactions (Sec.\ref{ssec:inter-values})
and then (Sec.\ref{sec:ladder-ham}) cast as many of them as
possible into effective LL description, now within the ladder
framework. Later in Sec.\ref{sec:RG} we study how the non-linear
interaction terms will affect the Luttinger liquid parameters and
what instabilities they can potentially produce. Finally in
Sec.\ref{sec:discus} we discuss our results for LL parameters in a
context of the experimental findings (Sec.\ref{ssec:disc-alpha}),
as well as the validity of the 1D approach itself in
Sec.\ref{ssec:disc-tperp}. We also discuss in
Sec.\ref{ssec:disc-disor} the role of substitutional disorder in
our model. The conclusions in Sec.\ref{sec:concl} close the paper.

\section{Band structure}\label{sec:bands}

\begin{figure}
  a) \includegraphics[width=\columnwidth]{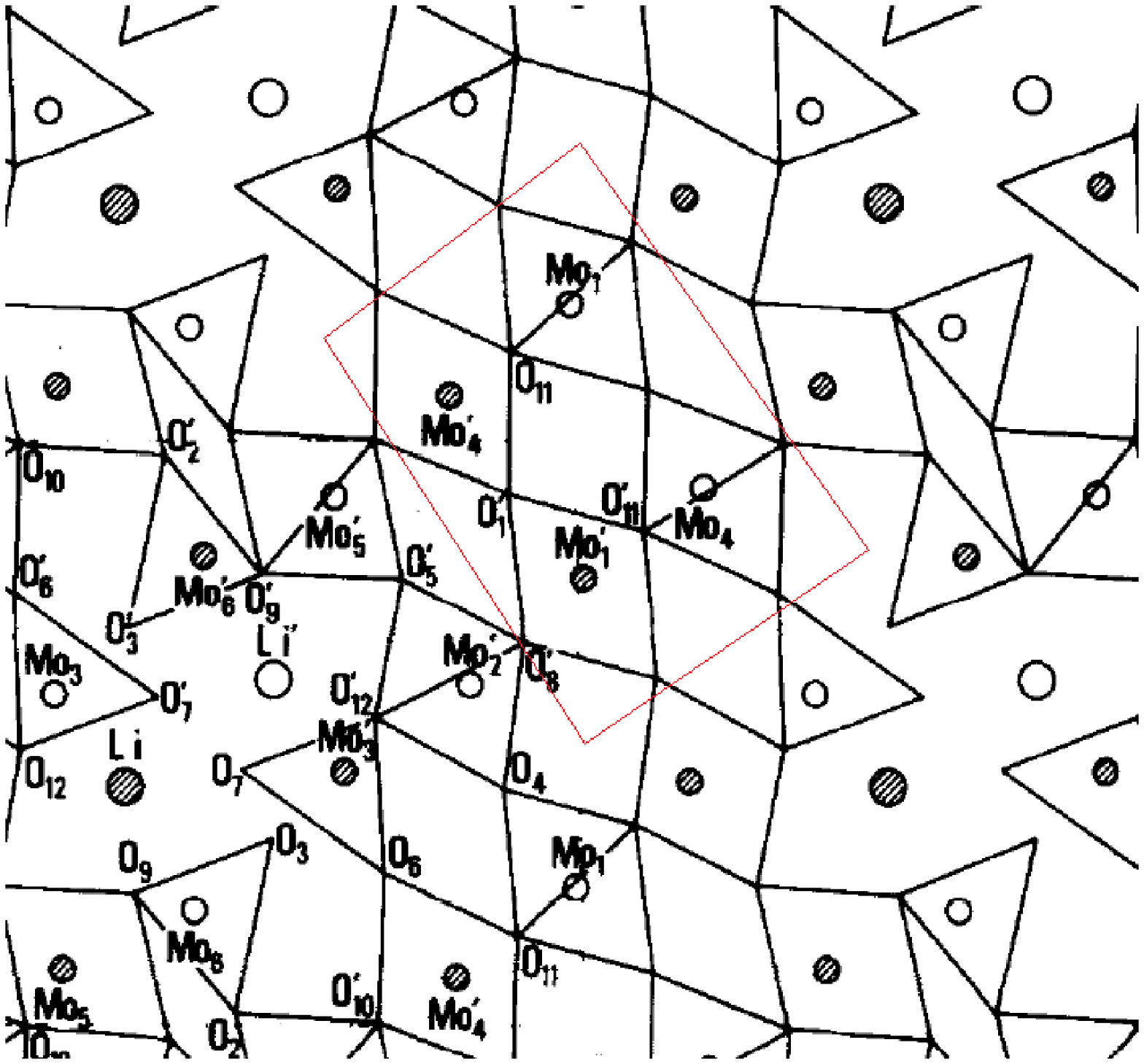}\\
  b) \includegraphics[width=\columnwidth]{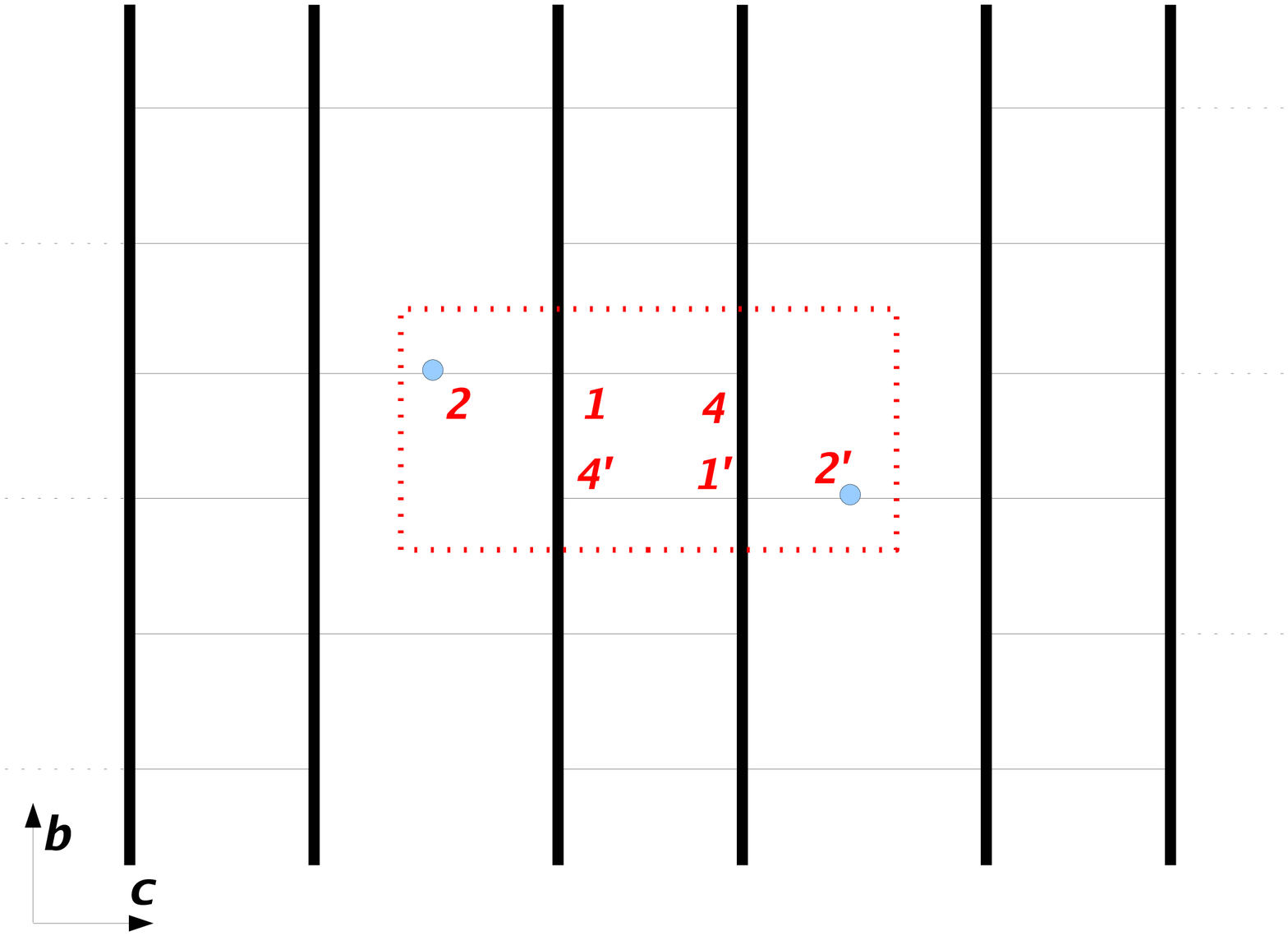}\\
  \caption{a) The crystal structure of purple bronze, after Ref.\cite{Onoda-structure-Xray}.
  A cut perpendicular to highly conductive b-axis is shown. The atoms which belongs to zig-zag chains are indicated.
  b) A simplified structure of the low energy model. We show the top view on b-c plane. The $t$ and $t_{\perp}$ hoppings
  as given in the first line of hamiltonian Eq.\ref{eq:ham-strong} are
  shown, sites 1 and 4 lie within the ladder, site 2 is outside it
  but $t_{\perp}$ hopping most likely goes through this site.}\label{fig:structure}
\end{figure}

The lattice space group and atomic positions within the structure
are known from experiment done by Onoda {\it et al}
\cite{Onoda-structure-Xray} recently confirmed by the neutron
diffraction experiment\cite{Santos-structure-neutrons}. This
rather complicated structure, which consist of well separated
slabs parallel to b-c plane, is presented on
Fig.\ref{fig:structure}a). Based on this crystallography knowledge
the electronic structure of Li$_2$Mo$_{12}$O$_{34}$ can be
calculated using density functional theory (DFT) \emph{e.g.} in
the local density approximation (LDA). Recent calculations
\cite{Thomas} lead to results which are globally consistent with
several previous computations\cite{Popovic-bron-DFT,
Canadell-DFT-old}. Overall the band dispersions agree well with
the measured results obtained by angle resolved photoemission
spectroscopy (ARPES) \cite{JWAllen-alphaRG}. For example it shows
a flattening of the two dispersive bands at about 0.4-0.5 eV below
$E_F$. Other bands are found at least 0.25 eV below $E_F$. The
only visible discrepancies are the relative vertical shifts of the
bands of order 0.1eV. The computed ratio between the Fermi
velocities along b- and c-crystal axis, is about 40, compatible
with the reported anisotropic 1D-like resistivity
\cite{Greenblatt-old-res-anis, Mandrus-optics,
Santos-transport-weird, Hussey-rhoB}. The velocity along $\vec{a}$
is even smaller. At low energies (around Fermi energy $E_F$) all
calculations gave qualitatively similar results. There are only
two bands which cross the Fermi energy and they are placed very
close to each other. They have a strong dispersion in the
$\Gamma-Y$ direction (b-axis) and barely no dispersion in the
perpendicular directions.

The two bands originate from a pair of zig-zag chains build out of
$Mo$ atoms inside $O$ octahedra. The low energies (close to $E_F$)
spectral function is located mostly on $4d$ (precisely $t_{2g}$)
orbitals of $Mo(1)$ and $Mo(4)$ atoms, using notation from
Fig.\ref{fig:structure}a). These two corners of zig-zag are
symmetrically inequivalent, thus dimerization is possible. However
their distinction comes more from out of chain, than in chain,
environment. The difference solely within a chain is hard to
notice. It is also hard to distinguish the corresponding gap on
the edge of Brillouin zone upon the analysis of the LDA band
structure (in the following we this assume that it is not larger
than $\sim 0.1 eV$).

The standard procedure (see Appendix \ref{app:tight-bind}) is to
fit the DFT band structure with an effective tight-binding model.
Along the b-axis (along the chains) the LDA calculations were done
in the reduced Brillouin zone (because of the above described
presence of two un-equivalent Mo sites). Then, in a tight binding
approximation, we expect that the band will be back-folded with a
gap at the boundary of a reduced Brillouin zone which corresponds
to the above mentioned strength of dimerization (not larger than
$\sim 0.1 eV$). In Li$_2$Mo$_{12}$O$_{34}$ case, while it is
rather easy to distinguish the lower half of dispersion bands
(however there are some peculiarities at $k^{c}\approx \pi$), the
upper bands hybridizes strongly with other bands (originating
mostly from oxygen orbitals). Their proper identification is then
difficult. The band gap seems to be quite small, however the
presence the hybridization makes the estimate quite difficult. In
the following we assume that it is $\leq 10\% t$.

Along the $c-$axis the dispersion relation is quite unusual (there
is a node at zero momenta along this axis) and this peculiar
feature appears within all independent DFT
calculations\cite{Popovic-bron-DFT, Canadell-DFT-old, Thomas}.
Here we are also dealing with the reduced Brillouin zone, but the
presence of the node means that the standard back-folding does not
apply at all. One has to try possible combinations of cosines(see
Appendix \ref{app:tight-bind}), keeping in mind two facts: the
unequal distances for intra- and inter-ladder hopping and the
large inter-ladder distance which implies that the next-nearest
neighbor hopping must be very small (only the light dashed lines
on Fig.\ref{fig:structure}a) could give non-zero contribution).

We were able to deduce (see Appendix \ref{app:tight-bind} for
details) that the hopping in the perpendicular direction
$t_{\perp}$ is $\sim 15meV$ (certainly smaller than $30meV$
($\approx 300K$)). One should also notice a large frustration,
change of sign, when hopping to next-nearest chain (interladder
hopping) with the amplitude of this hopping $\sim 10meV$. This
change of sign can be ascribed to a phase shift acquired when
hopping between pairs of chains, between two $Mo_1$ atoms via
$Mo_2$ octahedra (see Fig.\ref{fig:structure}a ).

Along the a-axis we have well separated slabs, with a void between
them filled by $Li$ atoms and Mo(6), Mo(3) tetrahedra. Electrons
residing there (if any) stay on energies few eVs away from $E_F$.
This explains why LDA give a very low dispersion in this
direction.

Significant deviations from LDA were seen in ARPES only below
0.2eV. Then the two bands seems to merge. This is also the energy
scale revealed by optics\cite{Mandrus-optics}: it shows a
formation of a first plasmon edge for electrons along b-axis
($\Gamma-Y$ direction) and a gap value in perpendicular direction.
Thus it sounds reasonable to take it as a point where the 1D
physics forms. This is the starting point (on the high energy
side) of the present study.

\section{Effective low energy hamiltonians}

\subsection{Tight-binding fermionic hamiltonian}\label{ssec:tight-bind}

The LDA-DFT results presented in Sec.~\ref{sec:bands} treat
electron-electron interactions at the mean field level. It is
assumed that a single electron moves in an average electrostatic
potential. However it is known that the higher order terms in
electron-electron interaction can bring very different physics,
particularly in the case of reduced dimensionality. Thus we need
to introduce them to our model.

The LDA results shows that a majority of carriers density at the
Fermi energy is localized on zig-zag chains formed by the Mo(1)
and Mo(4) octahedra (see Fig.~\ref{fig:structure}a)). We can
safely assume that the low energy physics is described by the
dynamics of these electrons. Then the low energy hamiltonian
formally writes:
\begin{multline}\label{eq:ham-tight-bind}
  H  = -t^{b}_{1}\sum_{ i\in even_b ,\sigma}c^\dagger_{i,\sigma}c_{i+\vec{b},\sigma} - t^{b}_{2}\sum_{ i\in odd_b ,\sigma}c^\dagger_{i,\sigma}c_{i+\vec{b},\sigma}\\
    - t^{c}_1 \sum_{ i\in even_c ,\sigma}c^\dagger_{i,\sigma}c_{i+\vec{c},\sigma} -t^{c}_2 \sum_{i\in odd_c ,\sigma}c^\dagger_{i,\sigma}c_{i+\vec{c},\sigma}\\
      - t^{a} \sum_{i ,\sigma}c^\dagger_{i,\sigma}c_{i+\vec{a},\sigma} + h.c.+\\
      \sum_{m, n ,\sigma, \sigma'}V_{nLDA}(r_{m}-r_{n})c^\dagger_{m,\sigma}c^\dagger_{n,\sigma'}c_{n,\sigma'}c_{m,\sigma}
\end{multline}

where the vectors $\vec{b}=[0,b,0]$, $\vec{c}=[0,0,c]$,
$\vec{a}=[a,0,0]$ define the $Li_{0.9}Mo_6 O_{19}$ crystal
lattice, summation runs over all ladder sites and depends on the
parity of a given site index along directions $\vec{b}$ and
$\vec{c}$. The $V_{nLDA}(r_{m}-r_{n})$ are the electron-electron
interactions not included in the (mean-field) DFT calculation and
the sum goes through positions of all carriers. The $t^{a-c}$ are
hopping parameters along each crystal axis, as estimated in the
previous section. As discussed there $t^{a}\approx 0$ (down to
$0.1meV$) so we can neglect it for the "high energy" range we are
interested in. The electrons are moving exclusively within one
slab (and mostly along the zig-zag chains). In
Eq.~\ref{eq:ham-tight-bind} we kept only nearest neighbor hopping,
since they are expected to be the dominant ones. The $t^{i}_{1,2}$
indicate a possibility of dimerization, namely different hoppings
along \emph{i-th} axis for even and odd bonds. Because Mo(1) and
Mo(4) are crystallographically different we included the
possibility that we are dealing with a dimerised chain,
half-filled in the reduced Brillouin zone. We denoted the two
corresponding hopping by $t^b_1$ and $t^b_2$. As explained in the
previous section we take $t^b_1 \approx t^b_2$ (with $10\%$
accuracy). The $t^{c}_{1,2}$ describe intra- and inter-ladder
hopping respectively (see Appendix \ref{app:tight-bind}). To
lighten the notations we will denote from now on $t^{b}$ by $t$
and $t^c$ by $t_{\perp}$, to emphasize that they correspond
respectively to hopping along and perpendicular to the chain
direction.

The resulting simplified tight-binding model is shown on
Fig.~\ref{fig:structure}b). We simplify the interactions in a similar way,
by explicitly considering the intra-chain and inter-chain parts of the interactions.
The microscopic Hamiltonian we take is given by
\begin{multline}\label{eq:ham-strong}
    H=-t\sum_{\langle i,j\rangle ,\sigma}c^\dagger_{i,\sigma}c_{j,\sigma} - t_{1\perp}\sum_{\langle i,j_1\rangle ,\sigma}c^\dagger_{i,\sigma}c_{j_1,\sigma} - t_{2\perp}\sum_{\langle i,j_2\rangle ,\sigma}c^\dagger_{i,\sigma}c_{j_2,\sigma}\\
      + U\sum_{i} n_{i\uparrow}n_{i\downarrow} + \sum_{m\neq n ,\sigma, \sigma'} V_{in}(r_{m}-r_{n})c^\dagger_{m,\sigma}c^\dagger_{n,\sigma'}c_{n,\sigma'}c_{m,\sigma}+\\
      + \sum_{m\neq n ,\sigma, \sigma'} V_{out}(r_{m}-r_{n}) c^\dagger_{m,\sigma}c^\dagger_{n,\sigma'}c_{n,\sigma'}c_{m,\sigma}
\end{multline}
where $t=0.8eV$ (see Table.\ref{tab:tight-bind}) is hopping
between nearest neighboring $i,j$ (denoted $\langle i,j \rangle$)
molybdenum atoms along zig-zag chains (b-axis) and $t_{1,2\perp}$
are a hoppings in a perpendicular direction, within a slab
(c-axis), between nearest-neighbors, $1$ is intra-ladder while $2$
is inter-ladder. The values of $t_{1,2\perp}$ are estimated in
Appendix \ref{app:tight-bind} and summarized in
Tab.\ref{tab:tight-bind}. As already mentioned $t_{\perp}< 20meV$,
so we can treat it as a perturbation on the top of the intra-chain
physics.

The strong correlations (last two lines in Eq.\ref{eq:ham-strong})
are usually parameterized by several quantities, which enter into
the effective hamiltonian: $U$ is the local on-site interaction
between charge densities of opposite spin
$n_{i\uparrow}=c^\dagger_{i\uparrow}c_{i\uparrow}$ (Hubbard term),
$V_{in}(r)$ is the interaction of two carriers placed inside the
same chain at a distance $r\neq 0$, $V_{out}(r)$ is the
interaction of two carriers placed in two different chains in a
distance $r$ (both $V_{in}(r)$ and $V_{out}(r)$ are defined
assuming an environment for which the LDA screening is included).
We will discuss their strength in Sec.~\ref{sec:strint}.

\subsection{Bosonic field theory of a 1D system}\label{ssec:LL-intro}

In the following sections, it will be shown that the interchain
couplings (both hopping and interactions) are reasonably weak
compared to the intra-chain ones. We can thus anticipate the need
to describe a strongly interacting one dimensional systems. In
such a case, a very convenient formalism to incorporate the strong
intra-chain interactions from the start is provided by the
Luttinger liquid formalism\cite{giamarchi_book_1d}. This formalism
allows to describe non-Fermi liquid state (Tomonaga-Luttinger
liquid (TLL)) that occurs in one dimension as a result of the
interactions.

The low energy dynamics of such a state is captured by collective,
bosonic modes representing charge and spin density fluctuations.
These fluctuations are connected to the two fields
$\phi_{\nu}(x)$, where $\nu = \rho$ for charge fluctuations and
$\nu = \sigma$ for spin fluctuations (linked with the fluctuations
of charge and spin density). These fields have two canonically
conjugate fields $\theta_{\nu}(x)$ linked with the respective
currents fluctuations. In terms of these collective variables the
hamiltonian reads\cite{giamarchi_book_1d}:
\begin{equation}\label{eq:LL-ham}
    H_{0}= \sum_{\nu} \int \frac{dx}{2\pi}[(u_{\nu}K_{\nu})(\pi \Pi_{\nu})^{2}+(\frac{u_{\nu}}{K_{\nu}})(\partial_{x} \phi_{\nu})^{2}]
\end{equation}
where $\Pi_{\nu}(x)=\frac{1}{\pi}\nabla\theta_{\nu}(x)$. All the
intra-chain interactions conserving the momentum are now only
fixing the precise values of the parameters $u_\nu$ (the
velocities of the corresponding charge or spin density modes) or
$K_\nu$ (the Luttinger parameters which control the decay of the
various correlation functions).

In particular the local spectral function behaves as a power law
$A(x=0,\omega,T=0)\sim \omega^\alpha$ with the following
(interaction dependent) exponent:
\begin{equation}\label{eq:alpha-chain}
    \alpha=\frac{(K_{\sigma}+K_{\sigma}^{-1}+K_{\rho}+K_{\rho}^{-1})}{4}-1
\end{equation}
For spin-rotationally invariant case $K_{\sigma} = 1$, thus
interactions, leading to the anomalous behavior, affects only the
charge sector.

We can thus take (\ref{eq:LL-ham}) as our starting point for the
chain physics, and study in the bosonized representation the
effects of the various inter-chain coupling terms. This will
ensure that at least the upper energy scales of the "high energy"
regime will be properly treated. The first step to determine the
Luttinger parameters $u_\nu$ and $K_\nu$ is to estimate the
strength of the various interactions, which we do in the following
section (Sec.\ref{sec:strint}). At energies lower than the Fermi
energy $E_F$ only two scattering channels are allowed $q\approx 0$
and $q\approx 2k_{F}$ (plus eventually higher harmonics). All
density-density interactions, with small $q$ momentum exchange,
can be incorporated into the hamiltonian given by
Eq.\ref{eq:LL-ham}. They contribute to a highly non-trivial
dependence of the TLL parameter $K_{\rho}$ (see
Sec.\ref{sec:LL-res}).

The remaining interaction terms produce non-solvable hamiltonians
of Sine-Gordon type, which in general is a functional $F[]$ of
cosine terms:
\begin{equation}\label{eq:LL-non-linear}
    H_{cos}= F[\cos(\sqrt{8}p\phi_{\nu}(x)),\cos(\sqrt{8}q\theta_{\nu}(x))]
\end{equation}
where $p,q$ indicate higher harmonics (scattering with larger
momenta exchange). Due to these terms the total problem allows
only for approximate solution (at least in terms of continuous
fields theory). The terms present in $H_{cos}$ are derived from
large momentum exchange interactions (of both intra- and inter-
chain origin). This is why their presence is analyzed carefully in
the further sections. They are usually treated using
Renormalization Group (RG) transformations, which allows to
extract the terms that affect the most the TLL physics. This
usually enables one to find the existence of gaps in the spectrum
of bosonic modes, sometimes even to predict the ground state phase
diagram. This kind of approach will be applied in
Sec.\ref{sec:RG}.

\section{Intra-chain physics}\label{sec:intra}

Let us first consider the physics taking place within a single
chain. We want to obtain (in Sec.\ref{sec:strint}) the values of
the various intra-chain interactions in (\ref{eq:ham-strong}) and
from there (in Sec.\ref{sec:LL-res}) the values of the Luttinger
parameters for the Hamiltonian (\ref{eq:LL-ham}).

\subsection{Strength of interactions} \label{sec:strint}

\subsubsection{U: local on-site interaction}

A fully self-consistent calculation has been performed in
\cite{Popovic-bron-DFT} and a value $U= 6.4$eV was found. This
value seems at first sight surprisingly large (for $4d$
electrons), but a recent study using constrained-RPA
\cite{Blugel-cRPA-U} allows to understand this result. Although
for bulk Mo $U\approx 3.8eV$, it was convincingly shown that the
suppression of the pure Coulomb value $U_0\approx 14eV$ is mostly
due to efficient screening in 3D of the d-electrons In our case
(as discussed in details below) the system is underscreened, which
entails also a significantly reduced plasmon frequency in
comparison with the pure Mo (from $\sim 15eV$ to $0.65eV$
\cite{Blugel-cRPA-U}). Thus the value $U=6.4eV$ is justified or
probably even modest. Then, if we take previously estimated
hopping $t$, we get  $U/t\approx 8$.

Note that this makes the local repulsion by far the largest energy scale in the problem.

\subsubsection{$V_{in}$: interaction inside a chain}\label{ssec:Vin}

In hamiltonian Eq.\ref{eq:ham-strong} we defined intra-chain
non-local interaction $V_{in}(r)$. In this subsection we will
discuss its strength both in real and in momentum space
$V_{in}(q)$. We are going to use shorthand notation $V_{in}\equiv
V_{in}(r=b/2)$, as this is usually the single parameter which
enters into so called $U-V$ model for a 1D chain. This parameter
is much more difficult to estimate than $V_{out}$ (see
Sec.\ref{ssec:inter-values}) as it involves the dynamics of the 1D
metal. Rather than trying to estimate it directly from the
interaction itself, we will simply adjust the value to be put in
(\ref{eq:ham-strong}) and (\ref{eq:LL-ham}) in order to reproduce
the experimental data on optical spectroscopy.

The idea\cite{Jacobsen-optic,Mila-UV-optic} is based on two
independent estimates of kinetic energy in the system. The first
one is related (by the optical sum rule) to the plasmon (edge)
frequency $I_p = \omega_p^{2}$ and gives us the total possible
kinetic energy available for the given number of carriers. The
same quantity can be computed as an integral $I_{\sigma}$ of the
optical conductivity $\sigma(\omega)$ (taken only over the highest
conducting band). The point is that the second estimate gives us
the real kinetic energy, renormalized by interactions.

Fortunately the necessary data is available for $Li_{0.9}MoO$. the
value of of the plasmon frequency can be read out from Fig.~2 in
Ref.\cite{Mandrus-optics} and gives $\omega_{p}=0.65eV$
($\omega_{p}$ extracted from the LDA \cite{Thomas} is even larger
$\approx 1.1eV$). The sum rule integral $I_{\sigma}$ was in fact
already evaluated in Fig.~3 of Ref.\cite{Mandrus-optics}.
Determining the value of the first saturation from the plot has of
course some error attached to it. We take $I_{\sigma}\approx 0.35$
($\pm 5\%$). This value is reasonable because at higher energies
one expects that the other bands start to intervene. The rest of
the procedure is straightforward, the ratio of bare and
interaction suppressed kinetic energy $I_{\sigma}/I_p$ equals to
$\approx 0.83$. Then the results of\cite{Mila-UV-optic}, together
with the previously estimated value of $U$ allows to predict
$V_{in}/t=1.2$ which means $V_{in}=0.95eV$ ($\pm >5\%$).

In addition to the interactions discussed up to now (which fit
within the $U-V$ model for a single chain) there can also exist
interactions between more distant neighbors (see
Fig.~\ref{fig:interactions}). These interactions are non-zero
because of the under-screening, a characteristic property of the
1D systems. When the separation is larger than that between
nearest neighbor we can approximate the interaction by a continuum
limit $V_{in}(r>b) = V_{\rm Coul}(r)/\epsilon(r)$ where
$b=5.523\AA$ is a lattice constant along the
b-axis\cite{Onoda-structure-Xray} and $\epsilon(r)$ is a real
space dielectric constant. In 2D and 3D systems the electrostatic
potential is usually well screened thus the Thomas-Fermi
approximation, already used in the DFT calculation, is sufficient.
In our problem this means that for the very long length scales $r
\gg a$ the potential is well screened, $V(r \gg a)=0$ (where
$a=12.762\AA$ is the largest lattice constant, along the a-axis,
perpendicular to the slabs planes), because then the network of
the zig-zag chains can be thought as a homogenous 3D bulk. For
shorter distances between the interacting electrons $r\approx a$
and then the distance between the neighboring chains does matter.
The relation $a =2.1b$ implies that there are a few non-zero
interaction terms $V_{in}(r)$ for $a \geq r>b$.

The screened potential inside 1D wire behaves approximately like
$\ln(1/r)$ \cite{giamarchi_book_1d}, same relation holds for the
screening induced by the presence of other wires
\cite{Alejandro-screen1D2D}. In total we approximate that
$V_{in}(q=0)$ will get renormalized by a factor
$(1+\ln(a/b))\approx 1.6$. This gives a reasonable estimate for
$V_{in}(q=0)$, however one has to be aware that for the value of
charge mode TLL parameter the whole q structure has its
importance.

The values of real and momentum space interactions are summarized
in tables Tab.\ref{tab:str-coupl-r} and Tab.\ref{tab:str-coupl-m}.

\subsection{Values of TLL parameter in the charge sector}\label{sec:LL-res}

Clearly the density-density interactions caused by $U$ and $V$ are
not small perturbations compared to the kinetic energy. In fact
they are the largest energy scale in the problem, larger than
hopping along the chains. Thus they shall strongly affect the
charge sector, they are strong enough to give rise to well
pronounced non-Fermi liquid properties. We thus need to estimate
the TLL parameters entering (\ref{eq:LL-ham}). As we will prove in
later sections the following hierarchy of energies holds $U\gg
V_{in}>t>V_{out}>J_{eff}$ (where $J_{eff}$ is an effective
superexchange which determines energy scales for spin sector).

We thus start by solving the TLL problem for a single chain and will consider
later the interchain couplings.

Even when $U\gg V_{in}>t$ we stay \cite{Haldane-1Dmain} in a
gapless phase within the TLL universality class, but computing the
values of the Luttinger liquid parameters $K_{\rho}$ beyond the
weak coupling limit is usually a very difficult problem
\cite{giamarchi_book_1d}, since one cannot compute them
perturbatively.

In the present case we have to deal with a quarter filled system
with a very small doping ($\delta$ around 2\%), and a small
dimerization (as we mentioned in Sec.\ref{ssec:tight-bind} it
cannot be larger than 10\% we take moderately 5\% in any future
calculations) and several non-zero interaction terms $V_{in}(r\geq
b/2)$. As a first approximation we decided to analyze the physics
of these zig-zag chains by looking at a quarter filled extended
Hubbard chain with a local interaction $U$ and a nearest neighbor
one $V$.

If $U$ is the larges energy scale in the problem then the charge
sector can be mapped onto a spinless chain of fermions (or an XXZ
spin chain) for which the exact TLL parameters are known
\cite{Ejima-EPL-num, giamarchi_book_1d}. For a certain value
$V_{in}$ we have
\begin{equation}\label{eq:KXXZ}
 K_\rho^{XXZ} = \frac{\pi}{4\arccos\left[-\frac{V_{in}}{V_{in}^{c}(U)}\right]}
\end{equation}
where $V_{in}^{c}(U)$ is the critical value of intra-chain nearest
neighbor interaction for a given value $U$. When $U \to \infty$
the mapping is exact and the critical value is known
$V_{in}^{c}(\infty)=2t$. In this case we get $K_\rho^{XXZ}=0.3$
which sets the lower limit for the TLL parameter. This is a rough
estimate, but suggests that for further analysis we should use
approaches which are valid for the gapless phase.

A more precise estimate can be obtained from a numerical study of
the extended Hubbard model, and an estimation of the TLL
parameters in the usual way via thermodynamic quantities
\cite{Mila-Zotos-num, Ejima-quart-dim-dop, Ejima-EPL-num}. From
the relevant plots one reads out that our model is located
somewhere in the range $K_\rho\in (0.3,0.37)$. One can try to give
an even better estimate using Eq.\ref{eq:KXXZ} even when U is not
infinite (but still with U much larger that any other energy scale
in the problem). It is commonly believed that the value
$V_{in}^{c}(U)$ extracted from the numerics can give quite good
approximation when substituted into (\ref{eq:KXXZ}). Estimates of
critical $V_{in}^{c}(U)$ are usually done with higher precision,
than for arbitrary $U,V$. In our case $U=8t$ implies
$V_{in}^{c}(U)=2.6t$ \cite{Ejima-EPL-num, Sano-num+RG} and
$V_{in}^{c}(U)=2.75t$ from older work by Mila and
Zotos\cite{Mila-Zotos-num}. This gives $K_\rho=0.330$ (with $V/V_c
=0.72$) or $K_\rho=0.340$. We have also followed the critical
scaling analysis proposed in \cite{Sano-num+RG} and found similar
value $K_\rho=0.328$.

The exact solutions e.g. Bethe ansatz are available for a few
special cases. If we include a (quite weak) dimerization and (very
small) doping then the problem is located far away from any
integrable model. However the influence of both these
perturbations are known. The dimerization is able to lower a bit
value of $K_\rho$, this effect can be present in particular in our
case when we are not far from $V_{in}^{c}$
\cite{Ejima-quart-dim-dop}. The doping has the opposite effect,
however when one is not very close to critical point
$K_\rho^c=1/4$ and close to commensurate case then the effect is
negligible.

To conclude: the knowledge about the $U,~V_{in}$ we have collected
above allowed us to give a several estimates based on
complementary numerical calculations of $U-V$ models. All these
estimates gives $K_{\rho}\in(0.3;0.36)$ (we allowed for 10\% error
in the $U,~V_{in}$ due to uncertainty of parameters). This is also
in agreement with a very recent study for a system with finite
range interaction\cite{Assaad-QMC}: $K_{\rho}\approx1/3$ when we
take similar value of on-site term and account for the presence of
$V_{in}(r)$ interactions up to four-five Mo sites. However in this
last paper interactions with an exponential character were
studied, which means that they decay much faster than those in our
problem.

Note that this $K_\rho$ value is reasonably close to the one
$K_\rho^* = 1/4$ that would lead to a quarter-filled Mott
insulator in the presence of an infinitesimal $V_{in}$. The system
will be thus very susceptible to the precise value of $V_{in}$.
Given the accuracy of our estimation we will take $K_{\rho}=1/3$
for further calculations. This value corresponds to the case for
which the $4k_{F}$ charge fluctuations decay with the same
exponent than the $2k_{F}$ charge and spin density
fluctuations\cite{giamarchi_book_1d}.

\subsection{The remaining interaction terms}\label{ssec:intra-rest}

By now we have studied interactions in real space (which we
believe can be useful for numerical studies), while cosine terms
beyond the TLL general expression (\ref{eq:LL-ham}) are defined in
the reciprocal space representation. These values of scattering
for large momenta exchange will be used as an input for the
discussion done in Sec.~\ref{sec:RG}.

The Hubbard $U$ interaction has a form of a delta function in real
space thus in momentum space it contributes equally to small and
large momenta exchange scattering (provided they are the
intra-chain ones). The previous considerations also imply that,
the Fourier transformed $V_{in}(q)$ is a weakly decaying function
(slower than logarithm), so it affects the amplitude of scattering
processes with large momenta exchange. As it was already expressed
above the intrachain interactions are much larger then $t$ thus it
is not straightforward to obtain the value of backscattering they
cause. In fact the $K_{\rho}=1/3$ parameter given above, known
from numerical studies of similar models, is a fixed point value
$K_{\rho}^*$ which means that it contains a part of these
contributions. Precisely it is the part included in the simplified
$U-V$ model. The longer range, slower decaying interaction leads
to smaller value of $K_{\rho}^*$. This suggests that the proper
value of $K_{\rho}^*$ in our problem which contains long range
interaction can be even smaller.

In our case the charge sector is renormalized by the possible
umklapp terms $g_3$, so the last statement can be rephrased: the
smaller final value of $K_{\rho}^*$ can be linked with larger
value of initial, bare $g_3^0$. The precise estimation of the
amplitude of $g_3^0$ is a very difficult task, the detail
discussion of the bare \emph{large-q}, intra-chain, term will be
done in Sec.~\ref{sec:RG}.

In addition, the $\epsilon(r)$ when computed using RPA (beyond
standard DFT) gives a well known phenomena, the Friedel
oscillations (in real space). The effect comes from the peculiar
screening (singular susceptibility) at large momenta $q=2k_F$. In
our case, due to the value of $k_F$ along chains, it gives an
extra gain of energy for electrons located every second site. This
affects the large momentum exchange part of interaction
$V_{in}(q\sim 4k_{F})$. The value of this gain can be estimated
using the fact that, for 3D metal, the additional oscillating part
$V(r)\sim r^{-3}$ thus we give an estimation $V_{\rm Frid}\leq
V_{in}(r=b)/8 = 0.1eV$. It is not as large as U or V, but can be
significant if we compare it with $t_{\perp}$.

The U-V model that we have considered so far (in particular in
Sec.\ref{sec:LL-res}) is of course only an approximation of the
intra-chain physics and interactions. It should capture most of
the effects at the energy scales we are considering, however it
may slightly underestimate the strength of interactions.

\subsection{Spin mode}\label{ssec:spin}

All the interactions considered up to now were connected with the
charge sector. In 1D systems, because of the spin-charge
separation, electrons spin degree of freedom should be discussed
separately. For a half filled chain the knowledge about $t$ and
$U$ allows to estimate the spin-spin exchange (superexchange)
constant $J=t \frac{4t}{U}$. This determines the energies at which
spin sector starts to play a role. The problem of purple bronze is
more complex since the compound is quarter filled.

For such cases the formula for superexchange interaction can be
still obtained from second order perturbation
theory\cite{Ejima-quart-dim-dop}. It reads:
\begin{equation}\label{eq:Jsuperex}
    J_{eff}=\frac{4t_{2}^2}{8t_{1}+2U+V_{in}-2\sqrt{(U-V_{in})^2+16t_{1}^2}}
\end{equation}
where we have taken into a account the fact that due to
dimerization there are two slightly different alternating hopping
$t_1$ and $t_2$ along the chain taking $t_2 - t_1 \approx 0.05 t$
and using the values of $U$ and $V_{in}$ obtained in
Sec.\ref{sec:strint} we obtain:
\begin{equation}
 J_{eff}=0.2eV
\end{equation}
It is indeed significantly smaller than $t$ (and also the  charge
sector interactions $U, V_{in}$) which implies that charge
dynamics will dominate the $0.2-0.02eV$ energy range. However it
is a non-negligible value in the sense that even for energies as
high as $\omega\sim 0.1eV$ (in the middle of the considered "high
energy" regime) the spin excitations are coherent and dispersion
is linear thus a TLL description with spin and charge modes is
applicable. The spin-incoherent TLL\cite{Fiete-RevModPhys-SILL} is
not an appropriate framework for our problem.

For a spin-rotational, SU(2) invariant model we have
$K_{\sigma}(T=0)=1$. For the interacting case the spin velocity
$u_{\sigma}$ is smaller than $v_{F}\approx 2t$, in particular
$v_{\sigma}(U\rightarrow\infty)\rightarrow J$. This last value is
comparable with the one observed in
experiment\cite{JWAllen-1D+2D}. The only candidate to break the
SU(2) symmetry would be spin-orbit coupling $D_{LS}$ on the
heaviest atom, molybdenum. A series of experimental and LDA
studies allow to set its value in bulk bcc Mo\cite{Iverson-Mo-LS}
to $D_{LS}=100meV$. This value was obtained for the $\Delta$ point
of a Brillouin zone in the Mo-bcc crystal. In our problem it
should be smaller because the active electrons have mostly
$t_{2g}$ character (with a larger \textbf{\emph{$l$-}}number
value). Thus $D_{LS}$ can be treated as a perturbation for $J$,
contrary to $U$ for $t$ in a charge sector. This implies that
$K_{\sigma}$ deviates from the non-interacting value
$K_{\sigma}=1$ much less than $K_{\rho}$. To be precise from weak
coupling theory (applicable in this case) we
know\cite{giamarchi_book_1d}:
\begin{equation}\label{eq:Ksigm-weak-coupl}
    K_{\sigma}=\sqrt{\frac{1+\frac{D_{LS}}{2\pi J}}{1-\frac{D_{LS}}{2\pi J}}}
\end{equation}
from which we predict $K_{\sigma}=1.1$ which can increase the
Green's function exponent $\alpha$ only by $0.01$. The spin sector
thus cannot be responsible for the experimentally observed values
of the $\alpha$ exponent (defined in Eq.\ref{eq:alpha-chain})
which are of order $\approx 0.5$. However the influence of the
spin-orbit coupling should be taken into account when one studies
the physics taking place lower energy scales, which are beyond the
scope of this paper.

\section{Physics between the chains}\label{sec:inter-phys}

After giving the description of a single chain physics we move
beyond this model and study the strength (and the role) of
interactions between carriers moving in different chains.

\subsection{$V_{out}$: the values of interactions between the
chains}\label{ssec:inter-values}

\begin{figure}
  \includegraphics[width=\columnwidth]{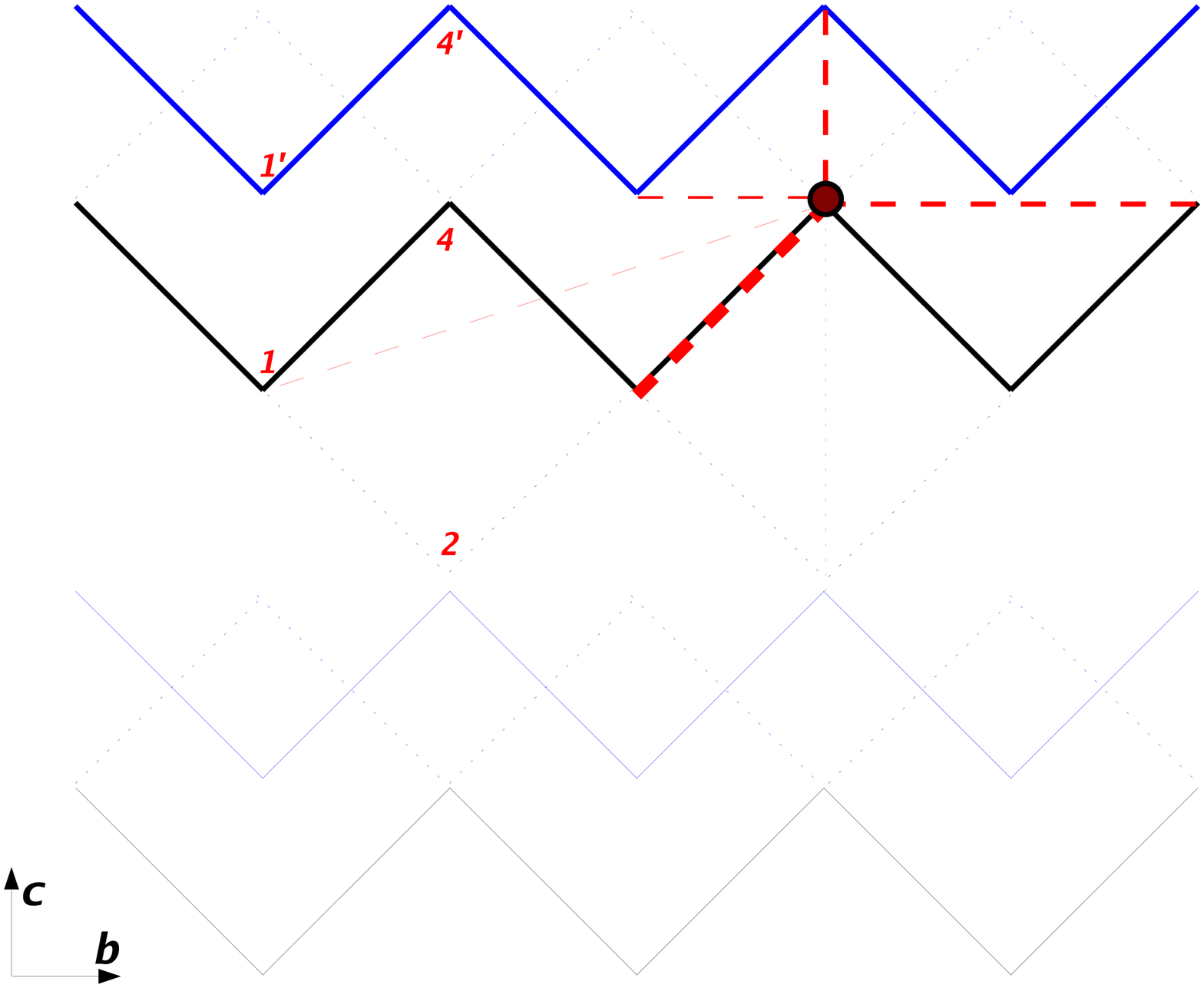}\\
  \caption{(color online) A real space image of strong interactions terms possible in purple bronze; the top view of a conducting slab is given.
  Interactions between an electron located on a dot-site and charges on different sites are indicated as red dashed lines of different
  thicknesses. Both $V_{in}$ (dashed lines within a black chain) and $V_{out}$ (dashed lines between the black and blue chains) are shown.
  The numbering of sites is like on  Fig.\ref{fig:structure}.}\label{fig:interactions}
\end{figure}

The density-density interactions can be estimated as a standard
Coulomb interaction between charges in a 3D dielectric, as it was
done in Ref.\cite{Popovic-bron-DFT}. Those authors gave estimate
for a value of interaction from standard electrostatic Coulomb law
(with the simplest static screening):
\begin{equation}\label{eq:VoutCoulomb}
    V_{out}(r=a)=\frac{1}{\kappa \epsilon_0 a}
\end{equation}
which in fact gives the value for two electrons in two different
2D slabs. We are more interested in interactions inside the slab
which is obviously larger because (the smallest) interchain
distance is then $c/2=0.4a$. A more fundamental problem is the
value of the dielectric permittivity $\kappa$. In the previous
work \cite{Popovic-bron-DFT} the bulk $\bar{\kappa}\approx 10$ was
used, which is typical for bulk semiconductors with similar value
of a gap for oxygen states($\Delta_{O}\approx 2eV$). Taking into
account the very weak metallic character along the c-axis the
semiconductor approximation is correct. For further neighbors,
with many oxygen atoms in between, one can take the bulk
$\bar{\kappa}$ value. However for the nearest chains, as there is
only one raw of oxygen atoms in the space between them, such large
$\kappa$ value overestimates the screening provided by sparse
environment\footnote{in a thin layers of a semiconductor $MoO_3$
(a material with locally similar structure) the measured value is
$\kappa=2.5$\cite{Kant-Mo-permit}}. It is then convenient to take
a function $\kappa(r)$ such that $\kappa(c/2)$ is reduced by
$30\%$ in comparison with $\bar{\kappa}$. For larger $r$ distances
at approximately $r\approx a$ it saturates to bulk value
$\bar{\kappa}$. With this set of values we estimate
$V_{out}=0.55eV$. For $r\rightarrow \infty$ the metallic character
along b-axis intervene and $\kappa(r) \rightarrow \infty$. Thus
the large distance interaction is strongly suppressed. In such a
case the above estimated value for $V_{out}$ can be taken as the
density-density interaction ($V_{out}(q \sim 0)$) between the
nearest chains. The above estimate was done between two nearest
chains which form a pair (see Fig.~\ref{fig:interactions}). On the
other side of each chain there is another neighbor which is placed
two times further. According to (\ref{eq:VoutCoulomb}) these
interactions $V_{out2}$ (we defined the inter-ladder term in
analogy with \emph{e.g.} $t_{c 2}$ in
Eq.\ref{eq:tight-bind-disp-def}) are at least two times smaller.
We then estimate (keeping $\kappa(r)$ in mind) $V_{out2}=0.2eV$.

\begin{figure}
  a)\\
  \includegraphics[width=0.9\columnwidth]{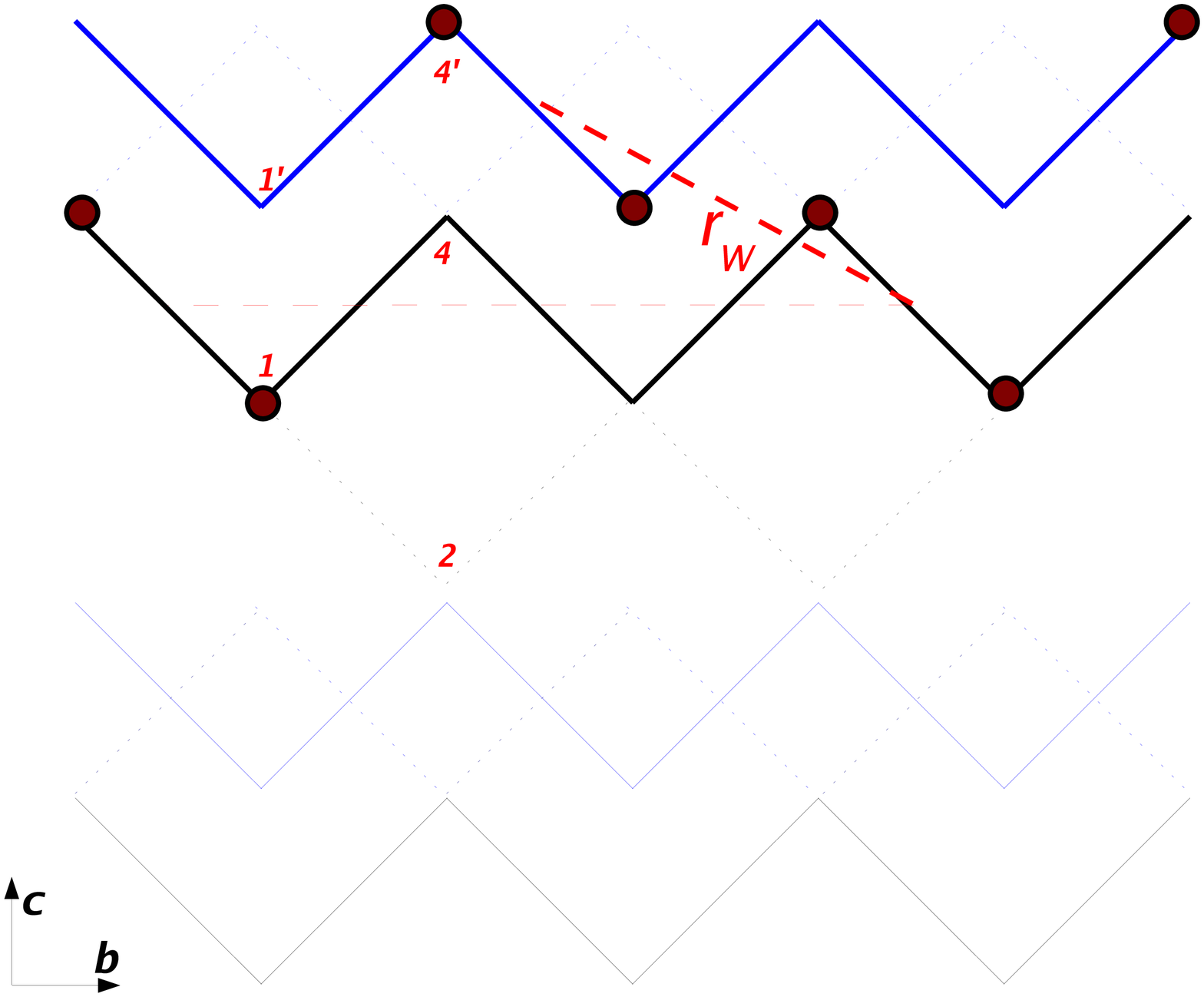}\\
  b)\\
  \includegraphics[width=0.9\columnwidth]{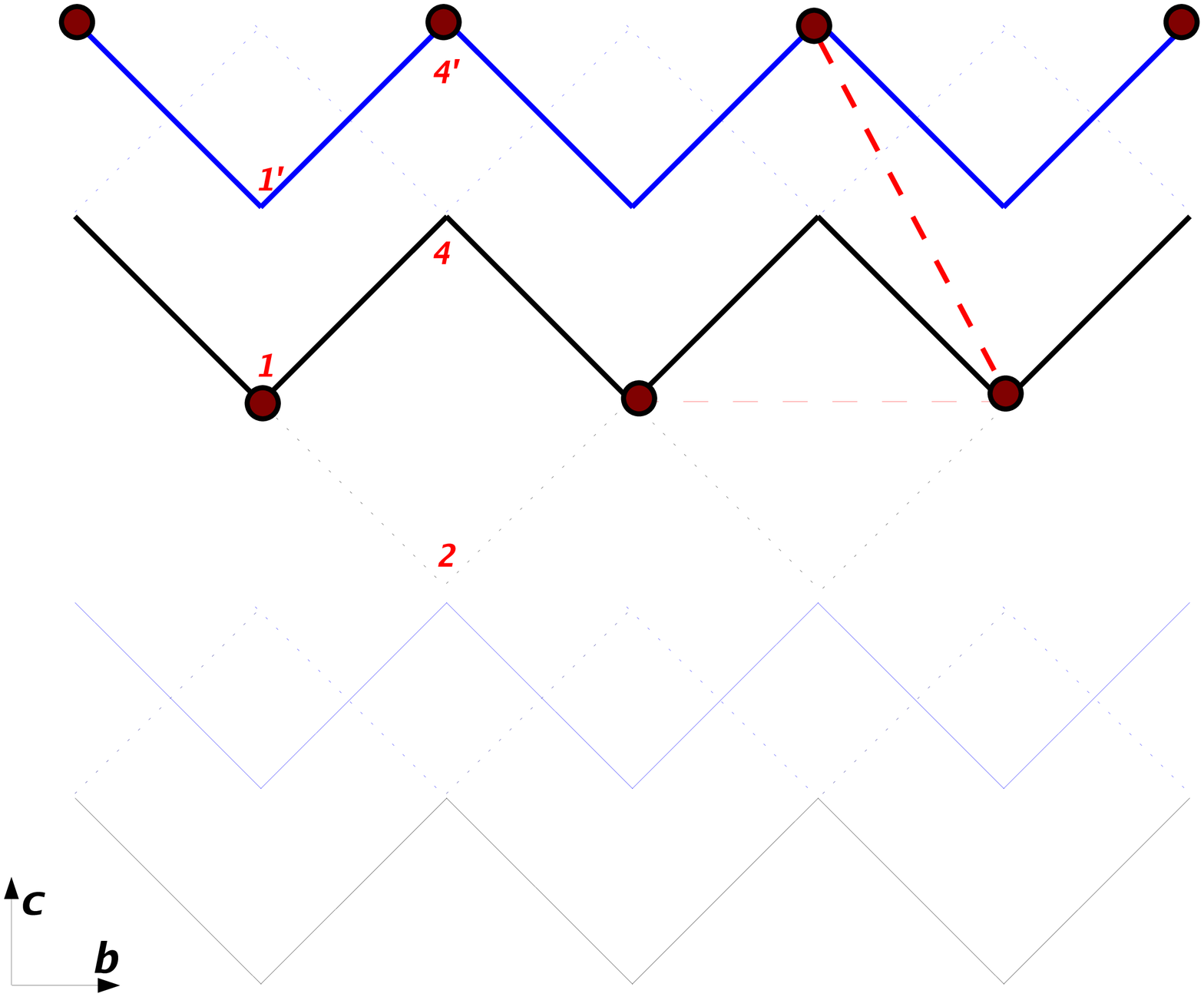}\\
  \caption{A schematic side view of a ladder structure (which enters to the low energy description of purple bronze)
  with: a) the $2k_{F}$ $\pi-$DW instability shown (The light, dashed line correspond to
  the distance $r_{\pi}$) b) the $4k_{F}$ $\pi-$DW instability shown.}\label{fig:out-instab}
\end{figure}

The treatment of terms with \emph{large momentum exchange} is
more complex. There is a $q=2k_{F}$ term which locks the
interchain density wave, the so called $\pi-CDW$, which is shown
on the Fig.~\ref{fig:out-instab}a) . From the figure it is clear
that the distances $r_{\pi}=\sqrt{(c/2)^2 + (2k_{F})^{-2}}$
between charges in such configuration are rather large (around
three times larger than the distance entering to previous
calculation for $V_{out}(q=0)$) thus the screening is quite
efficient and $\kappa[r_{\pi}]$ is definitely not reduced, but
probably even enhanced with respect to the bulk value. In fact not
only further chains, but also further slabs can intervene (because
$r_{\pi} > a$).

There will be either a very efficient screening (like in a metal),
or Coulomb potential approximation (but with enhanced $\kappa$) is
still applicable. We assume, optimistically, the second case and
use again (\ref{eq:VoutCoulomb}). Two ways of proceeding are
possible when one is interested in the staggered component of
the interaction between two chains. In real space (direct application
of (\ref{eq:VoutCoulomb})) one computes interactions with a
linear set of dipoles. Due to increase of $\kappa(r)$ the interaction (with further dipoles) decays rapidly, which
makes this straightforward approach quite tedious. In the
reciprocal space approach one must take into account the fact that
Fourier transform of $V_{out}(q)$ does decay in momentum space (it
will be a $1/q$ decay in a 2D case corresponding to a separate
slab). The second approach is simpler when one notice that
the previously computed $V_{out}=0.55eV$ correspond to $q=0\pm
(a/2)^{-1}$ (a distance between slabs sets the large distance
cut-off). Then the $1/q$ scaling allows to estimate that
$q=2k_{F}$ term will be suppressed by an extra factor four.
Overall we have:
\begin{itemize}
    \item a factor $1.5 \div 2$ from the value $\kappa(r_{\pi})$
    \item a factor $1.5 \div 3$ from the distance $r_{\pi}$
    \item a factor $\approx 4$ because we compute the staggered (large q)
    component
\end{itemize}
Taking all these factors into account we estimate that $2k_{F}$
component is reduced by at least one order of magnitude which
means $V_{out}(q=2k_{F})< 0.05eV$.

There is also the $q=4k_{F}$ term. Now the distance is smaller, thus
$\kappa$ is unaffected. However this term is strongly suppressed
due to above mentioned Coulomb character of $V_{out}(q)$. At short
distances corresponding to $r=(4k_{F})^{-1}$ the interaction may
even have a 3D character (with no screening) thus $V_{out}\sim
q^{2}$. By a similar reasoning like above we get
$V_{out}(q=4k_{F})\approx 0.05eV$. The smallness of inter-chain
large momenta terms (in comparison with density-density term)
implies that the local field corrections are small, so our mean
field model of screening is self-consistent. The influence of the
inter-chain exchange interactions (both $2k_{F}$ and $4k_{F}$) on
1D physics will be discussed in Sec.~\ref{sec:RG}.

To summarize the results for values of the strong coupling
parameters found in the last two section we present all parameters
in the tables Table~\ref{tab:str-coupl-r} and Table~\ref{tab:str-coupl-m}.
\begin{table}
  \centering
  \caption{The real space values of strong coupling parameters (given in eV).}\label{tab:str-coupl-r}
  \begin{tabular}{|c|c|c|c|c|c|}
    \hline
    $U$ & $V_{in}(r=b/2)$ & $V_{in}(r=b)$ & $V_{in}(r=2b)$ & $V_{out}(r=c)$ & $V_{out2}(r=2c)$ \\
    \hline
    6.4 & $0.95(\pm>5\%)$ & 0.5 & 0.2 & 0.55 & 0.2 \\
    \hline
  \end{tabular}
\end{table}

\begin{table}
  \centering
  \caption{The reciprocal space values of strong coupling parameters (given in eV).}\label{tab:str-coupl-m}
  \begin{tabular}{|c|c|c|c|c|}
    \hline
    $V_{in}(q=0)$ & $V_{in}(q=2k_{F})$ & $V_{out}(q=0)$ & $V_{out}(q=2k_{F})$ & $V_{out}(q=4k_{F})$ \\
    \hline
    1.6 & see Sec.\ref{ssec:umklapp} & 0.55 & $<0.05$ & 0.05 \\
    \hline
  \end{tabular}
\end{table}

\subsection{Luttinger liquid framework}\label{sec:ladder-ham}

The density-density interaction between the two neighboring chains
$V_{out}(q=0)$ can be included on non-perturbative level. Because
of the hierarchy of energies the presence of $V_{out}$ should be
included on the top of 1D chain $K_{\rho}$. The zig-zag chains are
grouped in pairs as shown on Fig.~\ref{fig:structure}b) . Inside
each pair we have two short links with interaction $V_{out}$ while
in between them there is only one link with much smaller
interaction $V_{out2}$. The ladder picture is then justified.

In the case of the ladder-like system (which means that there is
either much stronger intra-ladder hopping $t_{1\perp}\gg
t_{2\perp}$ or interaction) two more modes must be introduced. The
problem can then be expressed in two different possible basis. One
may work either in the chain basis with $\nu = \sigma 1, \sigma 2$
spin modes or in the total/transverse basis with symmetric and
antisymmetric spin modes $\nu= \sigma S, \sigma A$ (and
respectively the same for the charge sector). In general the
Greens function $\alpha$ exponent can be expressed as:
\begin{equation}\label{eq:alpha-chain}
    \alpha=\frac{\sum_{\nu}^{N}(K_{\nu}+K_{\nu}^{-1})}{2N}-1
\end{equation}
where $N$ is a number of modes. A significant $V_{out}(q=0)$ gives
a preference for symmetric and antisymmetric modes as it can be
diagonalized in this basis.

Thus within the ladder picture the two degenerate charge modes (in
the two chains of the ladder) split into two holons with two
different velocities. One of them (symmetric) describes the
fluctuations of the total charge density, while the other one
(anti-symmetric) describe the relative fluctuations in two
different zig-zag chains. Their values are:
\begin{equation}\label{eq:K_SA}
  K_{\rho S,A}=\frac{K_{\rho}}{1\mp K_{\rho} V_{out}/(\pi v_{F})}
\end{equation}
The respective values are: $K_{\rho S}\approx 0.30$ and $K_{\rho
A}\approx 0.38$. For energies well below $V_{out}$ the hamiltonian
of the charge sector is given as a sum of two bosonic modes plus a
non-linear part $H_{\rm cos}$ caused by large-q (exchange)
interactions:
\begin{multline}\label{eq:Hladd}
    H_{eff}=\sum_{\nu} \int \frac{dx}{2\pi}[(u_{\nu}K_{\nu})(\pi \Pi_{\nu})^{2}+(\frac{u_{\nu}}{K_{\nu}})(\partial_{x} \phi_{\nu})^{2}]\\ + H_{\rm cos}
\end{multline}
where as explained in Sec.\ref{ssec:LL-intro} now the summation
goes over four modes $\nu=\rho S, \rho A, \sigma S, \sigma A$. If
we neglect $H_{\rm cos}$ then the single-hole propagator (relevant
for ARPES), the density Greens function with chirality
$\vartheta$:
\begin{equation}\label{eq:Green-def}
    G_{\vartheta}^{<}(x,t)=\langle \psi_{\vartheta,\sigma}^{\dag}(x,t)\psi_{\vartheta,\sigma}(0,0)\rangle
\end{equation}

is known\cite{Orgad_A-LL} and can be written as a simple
generalization of the standard TLL result, as a product of four
gapless modes. For example for the right going hole:
\begin{equation}\label{eq:LL-propag}
    G_{R}^{<}(x,t)=\prod_{\nu}g_{\zeta_{\nu}}(x,t)
\end{equation}

where:
$$g_{\zeta_{\nu}}= \left[ (x-v_{\nu}t)\right]^{\zeta_{\nu}}\left[ (x+v_{\nu}t)\right]^{\zeta_{\nu}}$$

with the exponents $\zeta_{\nu}=(K_{\nu}+K_{\nu}^{-1})/8$. For a
numerical values of these exponents in our model and their
implications see Sec.\ref{ssec:disc-alpha}.

In the case when $H_{cos}$ is able to open a gap in a given mode,
the respective term in the product in Eq.\ref{eq:LL-propag} should
be substituted by a modified Bessel function of a second kind,
which gives the expected exponential decay of correlation
function.

The inter-ladder interaction $V_{out 2}$ are significantly smaller
and thus will not bring any novel physics at higher energy scales
(above 10meV), so this leave them out of the scope of this work.

\section{Renormalization Group study}\label{sec:RG}

\subsection{Statement of the problem with the inter-chain operators}

The interactions with small momentum exchange can be absorbed in a
definition of TLL parameters, but obviously there are also
scattering channels with the large momentum exchange. These
generate cosine type terms, which in principle can bring us away
from TLL universality class as defined in Eq.\ref{eq:LL-ham}.
There are several terms which are expressed as cosine operators in
the bosonization language. In addition to previously incorporated
umklapp terms there are the ones which emerge from the inter-chain
interactions. It is because usually they are the most pertinent
for the ladder system. They are:
\begin{description}
    \item[$q_1 =2k_{F}$] these are interactions originating from
    the presence of $V_{out}(q=2k_{F})$; they have a form
    $\cos(2\phi_{\rho A})\cos(\sqrt{2}\phi_{\sigma 1})\cos(\sqrt{2}\phi_{\sigma 2})$
    \item[$q_2 =4k_{F}$] these are interactions originating from
    the presence of $V_{out}(q=4k_{F})$; they have a form
    $g_{\perp}^{\pi}\cos(4\phi_{\rho A})$,
    and (only in the commensurate case)  the inter-chain umklapp $g_{\perp}^{u}\cos(4\phi_{\rho S})$
    \item[$t_{\perp}$] the single particle hopping induce several cosine operators
    \cite{Khveshchenko-RGtperp}, each of them in the form $\cos(2\theta_{\rho A})F[\cos(\sqrt{2}\phi_{\sigma 1}),\cos(\sqrt{2}\phi_{\sigma
    2})]$, where functional $F[]$ is a linear combination (there are also higher order hopping terms, but as they are proportional
    to $t_{\perp}^2$ we can safely neglect them)
\end{description}

The standard way to treat these terms is deriving, perturbatively
(usually on the single-loop level), the renormalization group (RG)
equations \cite{giamarchi_book_1d}. The RG equations allows us to
predict weather a given term will increase and affect the low
energy physics (become relevant) as the running energy variable
$\Lambda\approx max[\omega,T]$ decreases. To be precise the
quantity $l$ entering to RG equations is defined as $\Lambda \sim
\exp(-l)$ or to be precise $l=\ln(\Lambda/W)$, where $W=2t$. This
last formula allows to link two quantities (which we will do
throughout this section), however one should remember that while
$l$ is just a number, $\Lambda$ has an energy unit. In the
following we apply the following convention
$g_{\perp}^{\pi}[\Lambda_1]=V_{out}(q=4k_{F})/\bar{\Lambda}$ where
$\bar{\Lambda}=\pi v_F$.

The naive way would be to add RG equations describing these terms
into the previously determined TLL fixed point. Unfortunately, in
our particular problem this approach is not justified. Our problem
can be stated as follows: one to incorporate the above given
operators into the intra-chain, umklapp RG flow, which was already
accounted for.

We expect that the following physics takes place. Around $200 meV$
a gappless TLL appears and later undergoes the renormalization
flow. At this finite energy we know the values of interchain terms
caused by $V_{\perp}(q=4k_F)$ as they were estimated in the
previous section, but $g_3$ cannot be taken to be equal to zero
(it is not yet a fixed point as explained in
Sec.\ref{ssec:umklapp}). If we want to treat all instabilities
with $4k_F$ periodicity on equal footing then we have to begin the
flow with non-zero intra-chain $g_3$ and see how it will compete
or conspire with inter-chain instabilities. Estimating the value
of the initial $g_{3}$ which we need to substitute into RG
equations is a highly non-trivial task, our proposition on how to
tackle this problem is given in appendix \ref{app:bare-umkl}.

\subsection{The intra-chain, umklapp RG flow}\label{ssec:umklapp}

In Sec.\ref{sec:LL-res} we have given the values of intra-chain
TLL parameter $K_{\rho}$ within U-V model approximation. The point
is that these are already renormalized, the fixed point values
$K_{\rho}^*$, and as such already contain influence of intra-chain
umklapp scattering.

The fact that this operator is there and affects $K_{\rho}^*$
becomes clear from the following reasoning. As mentioned above, we
are working with weakly dimerized chains close to half-filling. In
the reduced Brillouin zone $k_{F}=0.487 \pi/b$ according to DFT
and $k_{F}=0.51 \pi/b$ according to ARPES. This discrepancy can be
understood. One should remember that the precise value of $k_{F}$
can vary as it depends on the relative value of chemical potential
inside a chain (determined also by strong correlations) with
respect to the local potential on the $Li$ ion (donor of
electron). In any case one must admit that the doping is low
enough, so in the "high energy regime" the zig-zag chain is not
able to recognize whether it is doped or not.

In addition an important remark has to be kept in mind: the above
form of umklapp operator is appropriate for a quarter filled
chain. This approach holds for weakly dimerized zig-zag chain, to
be precise for the case when amplitude of umklapp interaction is
larger than dimerization. From the analysis of the DFT band
structure we know that $t_1 - t_2 < 0.1eV$. After evaluating the
strength $g_3$ we also check if this assumption was consistent.

In this case the umklapp terms (so called $g_3$), in the i-th
chain, in the form $cos(\sqrt{8}\phi_{\rho i})$, has to be taken
into consideration. The RG equation for this instability reads:

\begin{equation}\label{eq:RG3}
    \frac{\partial g_{3}}{\partial l} = 8 g_{3}(1/4-K_{\rho})
\end{equation}

The critical value of TLL parameter for this operator is
$K_{\rho}^{c}=1/4$, so it seems to be irrelevant in our problem.
However in the case of Berezinskii-Kosterlitz-Thouless flow, like
the one described by eq.\ref{eq:RG3} there exists a straight line
on the $g_{3}-K_{\rho}$ plane, a separatrix between relevance and
irrelevance regime (gapped and critical phase). Essentially this
is because the compressibility of charge mode will also change:
\begin{equation}\label{eq:RG-Krho-umkl}
    \frac{\partial K_{\rho}}{\partial l}= -
    8 g_{3}^2 K_{\rho}^2 J_{0}(\delta^{eff}(l))
\end{equation}
and for large enough values of initial $g_{3}$, this equation
cannot be neglected, the RG flow cannot be assumed to be vertical
as we did before. Thus the amplitude of bare coupling $g_3$ can
play a role. As we will see below, in our problem the amplitude
$g_{3}$ does matter.

In the equation Eq.\ref{eq:RG-Krho-umkl} we kept the doping
dependence which is encoded inside the Bessel function of the
first kind $J_{0}(\delta(l))$. However below we assume that doping
is essentially equal to zero and we work in a commensurate case.
This assumption comes from the fact as we work in a constant
chemical potential (the chain is embedded in a whole crystal and
one knows the chemical potential of such a system from DFT
solution) the effective doping will get renormalized $\sim
-g_{3}^2 J_{0}(\delta^{eff}(l))$ already at higher energies.
Because of the same reason we neglect the renormalization of the
charge modes' velocity $\sim g_{3}^2 K_{\rho}
J_{2}(\delta^{eff}(l))$. From now on all we neglect all doping
dependence and set $\delta^{eff}=0$ in the "high energy regime".

\subsection{The RG analysis of the $4k_{F}$ terms}\label{ssec:4kF}

If the inter-chain $4k_{F}$ terms are considered alone then we
immediately see that they are less relevant than the $2k_{F}$
instabilities described in Sec.\ref{ssec:2kF} (it is because
$K_{\rho \nu}^{c}=1/2$). However there are several reasons why we
think that the $4k_{F}$ will be more important and decided to
investigate it first. The $4k_{F}$ depend only on charge modes,
which can be particularly important for energies larger than or
comparable with $J$. What is more, as we deduced before,
$K_{\rho}<1/3$ and then it is the $4k_{F}$ CDW instabilities which
are decaying slower. Finally there is an intra-chain umklapp term
with the same periodicity and large, bare amplitude.

Section Sec.\ref{ssec:umklapp} gave us necessary understanding of
the intra-chain RG flow, now we can proceed and introduce
inter-chain terms. The $q_2 =4k_{F}$ $\pi-wave$ scattering has the
following RG equation:
\begin{equation}\label{eq:RG2-p}
    \frac{\partial g_{\perp}^{\pi}}{\partial l} = 2(1-2K_{\rho
    A})g_{\perp}^{\pi}-g_{3}g_{\perp}^{u}
\end{equation}
and the inter-chain umklapp:
\begin{equation}\label{eq:RG2-u}
    \frac{\partial g_{\perp}^{u}}{\partial l} = 2(1-2K_{\rho
    S})g_{\perp}^{u}-g_{3}g_{\perp}^{\pi}
\end{equation}
In the above equations we have already introduced the so called
mixed terms caused by the presence of $g_3$ term (see below for an
explanation). We need to take the inter-chain terms together with
the standard umklapp $g_{3}$ to get a complete RG flow. These
terms will also affect the flow TLL parameters, the $K_{\rho
S,A}$, the symmetric/antisymmetric TLL parameters, which are
defined in different basis that the intra-chain $K_{\rho 1,2}$
ones (used in Eq.\ref{eq:RG3} and \ref{eq:RG-Krho-umkl}). The
intra-chain umklapp scattering may be rewritten in this basis
using the fact that the combination of two umklapps can be
expressed in a rather simple form:

\begin{multline}\label{eq:umklapp-comb}
g_3[\cos(2\sqrt{8}\phi_{\rho 1}) + \cos(2\sqrt{8}\phi_{\rho 2})]=\\
2g_3 \cos(4\phi_{\rho A}) \cdot \cos(4\phi_{\rho S})
\end{multline}

where we assumed that umklapp $g_3$ is identical in both chains 1
and 2. This can be justified by the crystal symmetry argument.
%Note that, inversely when one works in the intra-chain basis one
%may rewrite $V_{out}$ instabilities by analogy:

%\begin{multline}\label{eq:inter-comb}
%g_{\perp}^{\pi}\cos(4\phi_{\rho A}) + g_{\perp}^{u}\cos(4\phi_{\rho S})=\\
%2V_{out}(4k_{F})/v_{F}\cos(2\phi_{\rho 1})\cos(2\phi_{\rho 2})+...
%\end{multline}

%however here it is much less obvious that
%$g_{\perp}^{\pi}=g_{\perp}^{u}$, in fact due to effects of finite
%doping we rather expect $g_{\perp}^{\pi}>g_{\perp}^{u}$. In the
%intra-chain basis there would be also a large, off-diagonal
%$V_{out}(q=0)$ term which will compete with $g_3$. This two facts
%will give rise to basis rotation during RG procedure. The
%resulting RG would be extremely difficult to analyze, so we
%decided to work in the symmetric/asymmetric basis.

\begin{figure}
  \includegraphics[width=\columnwidth]{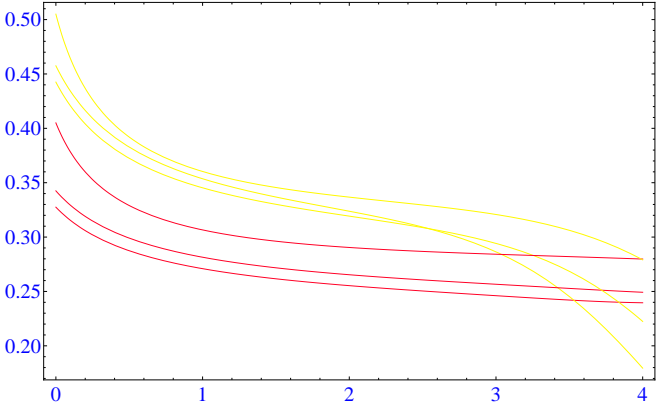}\\
  \caption{(color online) Three RG flows of $K_{\rho S}$ (red, three bottom curves at $l=0$)
  and $K_{\rho A}$ (yellow, three top curves at $l=0$) TLL parameters for different choices of initial parameters.
  The energy scale $\Lambda_2$ corresponds to $l\approx 3.5$.
  If all $K_{\nu}$ tend to be constant for increasing $l$, then the RG flow is so called "vertical".
  The initial values for three curves are (from top to bottom):
  (0.05, 0.04 , 0.5 , 0.5, 0.4);(0.05, 0.04 , 0.4 , 0.45, 0.35);(0.06, 0.05 , 0.38 , 0.45, 0.35);
  where the following notations are used $(g_{\perp}^{\pi}\bar{\Lambda}, g_{\perp}^{u}\bar{\Lambda}, g_3\bar{\Lambda}, K_{\rho A}, K_{\rho S})$.
  For the RG flow which give the bottom $K_{\rho S,A}$
  curves, the instabilities (not shown) are large $g_{i}\approx 1$ at
  the scale $\Lambda_2$ which means that the RG flow should be stopped. The other
  two flows do not suffer from this limitation. }\label{fig:K-flow}
\end{figure}

In the following we work with the perturbation described by
Eq.\ref{eq:umklapp-comb} in the symmetric/asymmetric basis. We
need to re-write the RG equations for the combined intra-chain
umklapp. Instead of Eq.\ref{eq:RG3} and Eq.\ref{eq:RG-Krho-umkl}
now we have:
\begin{equation}\label{eq:RG3-SA}
    \frac{\partial g_{3}}{\partial l} = g_{3}(2-4(K_{\rho S} + K_{\rho
    A}))+ 2 g_{\perp}^{u}g_{\perp}^{\pi}
\end{equation}
\begin{equation}\label{eq:RG-Krho-umkl-S}
    \frac{\partial K_{\rho S}}{\partial l}= - K_{\rho S}^2 \left(
    8 g_{3}^2 + (g_{\perp}^{u})^2 \right)
\end{equation}
\begin{equation}\label{eq:RG-Krho-umkl-A}
    \frac{\partial K_{\rho A}}{\partial l}= - K_{\rho A}^2 \left(
    8 g_{3}^2 + (g_{\perp}^{\pi})^2 \right)
\end{equation}
where (in the last two equations) we have distinguished the two RG
flows of symmetric and antisymmetric $K_{\rho}$s (instead of
single Eq.\ref{eq:RG-Krho-umkl}) and already included their
dependence on inter-chain interactions $V_{out}$. These equations
should be taken together with Eq.\ref{eq:RG2-p} and
Eq.\ref{eq:RG2-u} to obtain the full RG flow. Below we show in
Fig.\ref{fig:K-flow} the result of a direct integration of this
system of differential equations as well as a semi-quantitative
analysis of RG flow between energy scales corresponding to
$\Lambda_1 = 0.2eV$ (where TLL is likely to form) and $\Lambda_2
=0.02eV$. These are the limits of interest for the "high energy
regime" study. The physical reason of this limit will be given
below.

One can dispute whether a direct integration is valid when $g_3$
and $y_{\parallel}$ (see App.\ref{app:bare-umkl} for
$y_{\parallel}$ definition) are of order O(1). Such large terms
could in principle generate significant higher order operators
along RG. However there are no higher order terms proportional to
$g_3^n$ ($n\geq 2$). Then the higher orders mixed terms will be
either irrelevant (because of large $p$ value in functional
Eq.\ref{eq:LL-non-linear}) or of quite small amplitude.
Qualitatively the full RG flow can be thought as superposition of
two BKT flows, two hyperbolas on $g_i - K_{\nu}$ plane.

Let us first check what are the amplitudes different instabilities
$g_{i}$ at $\Lambda_2$, in the first order approximation. If we
neglect the flow of $K_{\nu}(l)$ then the solution of the
equations Eq.\ref{eq:RG2-p}, \ref{eq:RG2-u} \ref{eq:RG3-SA} is an
exponential function:
\begin{equation}\label{eq:RG-g-sol}
    g_{\nu}[2t*\exp(-l)]=g_{\nu}[\Lambda_1]\exp\left( \eta_{\nu}^* (-l-\ln(\Lambda_1/2t))\right)
\end{equation}
where $\eta_{\nu}$ is the dimension of the operator $g_{\nu}$,
\emph{e.g.} $\eta_{\perp}^{\pi^*}=2(1-2K_{\rho A}^*)$ The crucial
approximation we made in Eq.\ref{eq:RG-g-sol} is that we assumed
exactly vertical flows and took the fixed point values of
$K_{\nu}^*$, which tends to overestimate the strength of all
instabilities. As a result of Eq.\ref{eq:RG-g-sol} we get:
$g_{3}[\Lambda_2]\bar{\Lambda} \approx 0.1eV$,
$g_{\perp}^{\pi}[\Lambda_2]\bar{\Lambda}\approx 0.075eV$,
$g_{\perp}^{u}[\Lambda_2]\bar{\Lambda}\approx 0.08eV$. Note that
because initially $K_{\rho S}[\Lambda_1]<K_{\rho A}[\Lambda_1]$,
the inter-chain umklapp $g_{\perp}^{u}$ renormalizes more than the
$\pi-$DW one $g_{\perp}^{\pi}$, however one has to remember that
its bare amplitude is smaller due to doping effects. Already this
simplified reasoning shows that the amplitude of the irrelevant
term even at $\Lambda_2$ is still larger than amplitudes of the
relevant terms. This is a peculiarity of our problem. The presence
of a mixed terms will in fact enhance this property: because
$g_{\perp}^{u}\approx g_{\perp}^{\pi} < g_{3}$ then one can
interpret Eq.\ref{eq:RG2-p} and Eq.\ref{eq:RG2-u} as if $K_{\rho
S,A}^{c}$ was shifted from $1/2$ downwards. On the other hand in
Eq.\ref{eq:RG3-SA} we see that inter-chain instabilities push
(weakly) the RG flow of $g_3$ towards its separatrix.

Now let us move to the analysis of the $K_{\rho S,A}$ RG flows.
From Eq.\ref{eq:RG-Krho-umkl-S} and Eq.\ref{eq:RG-Krho-umkl-A} we
immediately realize that intra and inter-chain terms support each
other in lowering the values $K_{\rho S,A}$. Taking into account
the initial (at $\Lambda_1$) hierarchy of energies we can be sure
that the initial flow of $K_{\rho \nu}\sim g_i^2$ will be
dominated by $g_3^2$ while the reasoning of the previous paragraph
showed that around $\Lambda_2$ both terms are equally important.
Initially the flow slows down because an irrelevant $g_3$
decreases, but later it can speed up again due to relevant
inter-chain instabilities as seen on figure Fig.\ref{fig:K-flow}
where the result of numerically solving of
Eq.\ref{eq:RG-Krho-umkl-S}and Eq.\ref{eq:RG-Krho-umkl-A} is shown.
Estimating quantitatively $K_{\rho S,A}[\Lambda_2]$ can be
achieved in two ways.

\begin{figure}
  \includegraphics[width=\columnwidth]{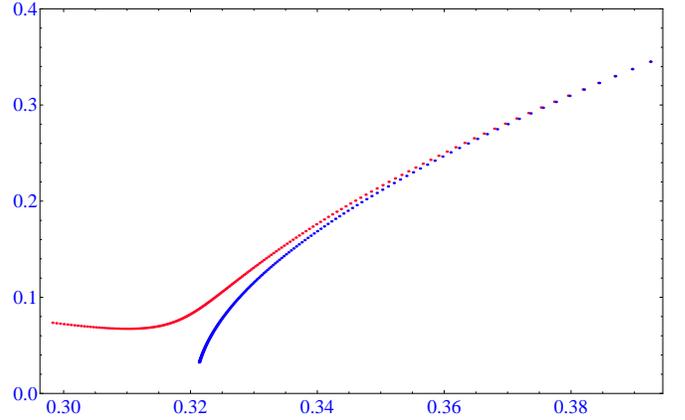}\\
  \caption{(color online) A comparison between BKT flow caused by a non-zero $g_3$ only (blue, bottom curve) and the
  result of our full RG flow (red, top curve). This is a parametric plot with $g_3[l]$ on the vertical axis
  and an average $(K_{\rho S}+K_{\rho A})/2$ on the horizontal axis. The initial values are:
  for blue line (0.0, 0.0 , 0.35 , 0.45, 0.35) and for red (0.06, 0.05 , 0.35 , 0.45, 0.35) with the same notation like in Fig.\ref{fig:K-flow}.}\label{fig:BKT-dev}
\end{figure}

First way (\textbf{i}) assumes the independence of intra- and
interchain BKT RG flows. We already know the influence of $g_3$
term alone, now we want to compute how much the
$K_{\nu}[\Lambda_2]$ would be lowered during the BKT RG flow
caused only by the inter-chain term, \emph{e.g.} if we keep only
$g_{\perp}^{\pi}$ term in Eq.\ref{eq:RG-Krho-umkl-A} for $K_{\rho
A}[\Lambda_2]$. We can use the procedure very similar to the one
applied in App.\ref{app:bare-umkl}, just that now we are moving
towards $l\rightarrow\infty$ along the RG trajectory. By analogy
with App.\ref{app:bare-umkl} reasoning we define the flow
invariant $A_{\perp}$. In the case of this RG flow $g_{\perp}^0
\approx 0.01 \ll 1$ thus $A_{\perp}\approx
(y^{\perp})_{\parallel}^0$, where $(y^{\perp})_{\parallel}^0=0.11$
is the distance of the bare $K_{\rho A}^0=0.39$ to $K_{\rho
A}^c=1/2$. We then obtain (from an analog of
Eq.\ref{eq:RG-itegr1}) that
$y^{\perp}_{\parallel}[\Lambda_2]=0.15$, which means that RG flow
caused $\Delta y^{\perp}_{\parallel}[\Lambda_2]=
y^{\perp}_{\parallel}[\Lambda_2]-A_{\perp}\approx 0.05$ of the
$K_{\rho A}[\Lambda_2]$ decrease \footnote{Because
$A_{\perp}\approx y_{\perp}^{0}$, one needs to take Taylor
expansion of $\tanh$ function in Eq.\ref{eq:RG-itegr1}.}. This
additional change of $K_{\rho A}[\Lambda_2]$ should be added on
the top of the value previously found from the U-V model (see
Sec.\ref{sec:LL-res}) $K_{\rho A}^*\approx 1/3$. However, because
of the $g_3$ influence, the $A_{\perp}$ is not a constant but it
increases during the flow so the above value of the $\Delta
y^{\perp}_{\parallel}[\Lambda_2]$ is underestimated. Such a
deviation from a single BKT flow, particularly pertinent at lowest
energies, is clear from figure Fig.\ref{fig:BKT-dev}.

The second way (\textbf{ii}) comes from the fact that, as pointed
out above, the $g_3$ term dominates most of the RG flow between
$\Lambda_1$ and $\Lambda_2$. Let us assume that the inter-chain
terms effectively adds up to the initial amplitude of
$g_3[\Lambda_1]$ (see Eq.\ref{eq:umklapp-comb}):
\begin{equation}\label{eq:g3-eff}
    g_3^{eff}[\Lambda_1]\approx
    g_3[\Lambda_1]+A_{aux}*2*V_{out}(4k_{F})/\bar{\Lambda}
\end{equation}
where the origin of $A_{aux}=0.6$ factor is explained in the end
of this section. Taking into account the results of
App.\ref{app:bare-umkl} we see that with this new effective
amplitude of the umklapp the initial point of the RG is located
quite close to the separatrix of intra-chain flow. The resulting
value of $K_{\rho \nu}^*$ is then going to be close to $K_{\rho
\nu}^c = 1/4$. This obviously overestimates the change of $K_{\rho
\nu}[\Lambda_2]$ because effectively in Eq.\ref{eq:RG-Krho-umkl-S}
we took a square of a sum instead of a sum of squares. There is
also a hidden assumption here that $K_{\rho A}\approx K_{\rho S}$,
but looking at figure Fig.\ref{fig:K-flow} we see that this works
well for all cases.

From the reasonings (\textbf{i}) (upper limit) and (\textbf{ii})
(lower limit) we conclude that within our approximation $K_{\rho
S,A}\in (0.25, 0.29)$. This is in agreement with the results
presented on Fig.\ref{fig:K-flow} and Fig.\ref{fig:BKT-dev}.

Finally we can try to estimate the value of a gap in the holon
spectrum. We choose to work with $g_{\perp}^{\pi}$ as it has the
strongest tendency to open a gap. We estimate the gap in the case
if this instability was acting on its own when $\Delta\approx W
\exp(-l^*)$ where $g_{\perp}^{\pi}[W \exp(-l^*)]=1$. For the
specific cases the $l^*$ is known\cite{giamarchi_book_1d}  for
example deep inside the gapped phase (self-consistent harmonic
approximation) or close to the separatrix of RG. Unfortunately our
problem does not belong to any of these, because $A_{\perp}\approx
0.12$ which for $\Lambda_2$ energy scale is neither very small nor
very large. We approximate the flow of the considered
$g_{\perp}^{\pi}$ by BKT flow which can be integrated out (see
also App.\ref{app:bare-umkl}):
\begin{equation}\label{eq:BKT-integr}
    \arctan\left(\frac{y_{\parallel}^{0}}{A_{\perp}}\right)-\arctan\left(\frac{y_{\parallel}}{A_{\perp}}\right)=A_{\perp}l
\end{equation}
If we use the fact that $y_{\parallel}^{0}\approx A_{\perp}$ and
$y_{\parallel}[2t\exp(-l^*)]\gg A_{\perp}$ (we are not far from
the separatrix) then we get $A_{\perp}l^* \approx \pi/4$. In fact
this value of $l^*$ is quite close to the ones we get from
numerical integration of the full flow. Then the gap is:
\begin{equation}\label{eq:gap-value}
    \Delta_{\perp}^{\pi}=2t\exp\left[-\pi/(4A_{\perp})\right]
\end{equation}
which gives $\approx 2meV$. The low initial value of the
$V_{out}(q=4k_{F})$ translates into quite small expected value of
the gap in the spectrum.

%when we are inter-chain terms we are deep in the gapped regime
%(and initial values of $V_{out}(q=4k_{F})$ are small) so one can
%use the to get a value of a gap[book]:

%$$\Delta_{\perp}^{\nu}=t(V_{out}(q=4k_{F})/t)^{1/\eta_{\perp}^{\nu}}$$

%where $\eta_{\perp}^{\nu}=2(1-2K_{\rho \nu}^{*})\approx 0.66$ is
%the scaling dimension of the considered $4k_{F}$ operators either
%is symmetric ($\nu=S$) or antisymmetric ($\nu=A$) mode. To obtain
%$\eta_{\perp}^{\nu}$ one should use the fixed point values of
%$K_{\rho \nu}^{*}$ estimated above. Substituting values from the
%previous section we get: $\Delta_{q2}^{S,A} \approx 10meV$.
%The relation $K_{\rho S} < K_{\rho A}$ implies that the symmetric mode
%has stronger tendency to be locked, on the other hand the bare
%amplitude of $g_{\perp}^{u^0}$ can be slightly lower.

The lowest energy (below $\Lambda_2$) flow of umklapp processes
are seriously affected by doping. If the $g_{\perp}^{\pi}$ is able
to open a gap before that then these other RG equations will be
changed by the presence of the gapped mode. For the intra-chain
umklapp instability the initial point lies very close (esp. if we
include terms beyond U-V model) to separatrix, which means that it
is very susceptible to these modifications. %If it happens that the effective
%$g_3^{eff}[\Lambda_1]$ as given by Eq.\ref{eq:g3-eff} lies above,
%then one should estimate a gap from the following formula [book]:

%$$\Delta_{3}= g_{3}^{eff}[\Lambda_1]V_{F} \exp(-\pi/\sqrt{2g_{3}^{eff}[\Lambda_1] (g_{3}^{eff}[\Lambda_1]- g_{3}^{c}[\Lambda_1])})$$

%However we know that certainly the distance from the separatrix is
%extremely small, assuming that it  $< 1\% g_3^0$ then equation
%above gives as a result $\Delta_{3}\approx 1meV$. This implies
%that a gap caused by intra-chain umklapp cannot be larger than
%this value, a quantity of the same order like the the $4k_{F}$
%gap. The above reasoning is not able to reveal which is the
%dominant instability.

Usually in such a case one thinks about two competing
instabilities taking place in two different basis (intra- and
inter-chain). Our problem is different, the cooperation of
instabilities is realized. This becomes clear if one assumes that
he tends to be locked at minimum \emph{e.g.} corresponding to
$\sqrt{8}\phi_{\rho 1,2}=\pi$ (then $\cos(\sqrt{8}\phi_{\rho
1,2})=-1$). For this specific value of $\phi_{\rho 1,2}^{0}$ the
combined inter-chain terms (see Eq.\ref{eq:umklapp-comb}) give
non-zero and negative value ($A_{aux}\approx -0.6$). This
corresponds to an additional energy gain caused by an auxiliary
instability. It is a rare case when the two gaps does not exclude
but can help each other. In particular this validates the
assumption made in Eq.\ref{eq:g3-eff}.

\subsection{The analysis of the $2k_{F}$ terms}\label{ssec:2kF}

The RG equation for the $2k_{F}$ backscattering operator
reads\cite{giamarchi_book_1d}:

\begin{equation}\label{eq:RG1}
    \frac{\partial g_{\pi}^{2k_{F}}}{\partial l} = (2-K_{\rho A}-K_{\sigma A})g_{\pi}^{2k_{F}}
\end{equation}

It is the most relevant instability (already for $K_{\rho A}<1$)
if one assumes SU(2) symmetric case in the spin sector when
$K_{\sigma A}=1$, so naively we would expect it to dominate the
low energy physics. However there are details that matter. First,
the initial amplitude of this term is quite low. Second, the RG
flow of this term will be strongly perturbed:
\begin{itemize}
    \item at high energies ($\Lambda \approx 0.2eV$) the spin sector is still in the incoherent
    regime, certainly the dispersion is not yet linear (mind the value of $J_{eff}$). Thus all
    terms which contain spin dynamics, like $\cos(\phi_{\sigma})$ will strongly fluctuate
    \item the $2k_F$ terms has to compete with $4k_F$ instabilities described in the previous section:
    large $g_3$ (in the initial part of the flow) and relevant
    $g_{\perp}^{\pi}$ and $g_{\perp}^{u}$ in the lower energies
    \item at lower energies ($\approx 0.015eV$) several terms generated by the perpendicular hopping ($\sim t_{\perp}$) will
    start to intervene. Although they are irrelevant (connected with the field $\theta_{\rho A}$ \cite{Khveshchenko-RGtperp}),
    still their initial amplitude is significantly (three times) larger than $V_{out}(q=2k_{F})$. We then expect that physics will be
    dominated by a competition between these two types of cosine
    terms, as it was studied in Ref.\cite{Suzumura-t-perpRG2, Suzumura-t-perpRG} where the term
    confinement-deconfinement transition was coined to describe
    the physics.
\end{itemize}
Based on results from Ref.\cite{Suzumura-t-perpRG2,
Suzumura-t-perpRG}, one may expect that due to last mechanism the
$2k_{F}$ instability is weakened. In any case because $K_{\rho A}
\ll K_{\rho A}^{c}=1$ the $2k_F-$RG flow is vertical. This implies
that $K_{\rho A}$ value (and thus also Green's function $\alpha$
exponent) is very weakly affected by the presence of the $2k_F$
terms.

\section{Discussion}\label{sec:discus}

\subsection{Comparison with experiment}\label{ssec:disc-alpha}

In this last part we wish to compare the estimated Luttinger
liquid parameters with experimental findings. Experiments look at
low energy ($l>1$), long distance behavior ($r> k_{F}^{-1}$),
which we should keep in mind for the rest of this section. The
crucial question is: what is the value of TLL parameters $K_{\nu}$
that enters into measured correlation functions? Obviously the
answer cannot depend on the point where we arbitrarily decide to
stop the RG flow, one has to keep in mind that $g_{i}$ terms
usually are still finite at such a point. As discussed in
Ref.\cite{Schulz-corr-fun}, the correlation functions are of the
form:
\begin{equation}\label{eq:corr-fun-power}
    R(r)=\left(\frac{r}{W}\right)^{K^{c}}\exp\left(\int_{0}^{r/W}y_{\parallel}[l]dl\right)
\end{equation}
where $|r|=\sqrt{x^2+(v_{\nu}\tau)^{2}}$ is a distance in a
time-space domain, W is an ultraviolet cut-off of the problem
($\sim 2t$), $y_{\parallel}[l]=K_{\rho}-K_{\rho}^{c}$ is a
deviation of TLL parameter from the critical value $K_{\rho}^{c}$
of the considered flow. Thus the observed value could be
interpreted not as $K[l]$, but rather as weighted average of
$K[l]$ over longer and longer length-scales. In particular if we
flow to a gapless phase then the observed $K_{\nu}$ corresponds to
$K_{\nu}^*$, which is then equal to the invariant of the flow $A$.

Our problem is particularly difficult, because the answer to the
above question strongly depends on yet unknown lowest energy
physics. Indeed in our reasoning we have neglected several effects
($t_{\perp}$, $D_{LS}(q=2k_{F})$, proximity to Mott insulator,
disorder) whose amplitudes are certainly smaller than
$\Lambda_{2}\approx 20meV$ and which seriously harm TLL below this
energy scale. Based on this we can reassure the validity of our
approach only down to $\sim 20meV$.

Basically there are two distinct behavior which can occur at these
lowest energy scales. First, one of the above mentioned
perturbations can become pertinent just below the $\Lambda_2$
energy scale and thus stop the RG flow. In fact all these
perturbation in one way or another would not support further
lowering of $K_{\rho}$ value, thus the further $K_{\rho}[l]$ can
be then taken as vertical. The observed TLL parameter should then
be close to $K_{\rho}[\Lambda_2]$. The second possibility, which
is not unlikely, is that 1D RG will remain valid down to much
lower energies $\sim 1meV$ (see Sec.\ref{ssec:disc-tperp}). Then
as indicated in Fig.\ref{fig:BKT-dev} the RG trajectory always
stays close to separatrix. This statement is in fact quantified by
the values of invariants of the intra- and inter-chain RG flows:
$A_3 \approx A_{\perp}$, so it is always either $g_3[l]$ or
$g_{\perp}[l]$ keeping us close to separatrix. Then
Eq.\ref{eq:corr-fun-power} gives a power low with a
$K_{rho}^{c}=1/4$ exponent times a logarithmic corrections in the
form $\log(r/W)$.

%The answer is more complex in the case of gapped phase, however if
%one works at energies above the gap $\omega, T > \Delta$ then
%observed $K_{\nu}$ corresponds to $K_{\nu}[\Delta]$. As we will
%see below roughly one can take $\Delta\approx\Lambda_{2}$ This justifies the choice of
%the lower energy limit studied in this paper.

We are able to compare our results only with the experimental data
corresponding to the `high energy' range $\Lambda \in (0.02;
0.2)eV$. In fact this is also the energy scale down to which, with
no experimental doubt, the TLL persist. One should also keep in
mind that the TLL $\alpha$ exponent should be computed within the
ladder model introduced in Sec.\ref{sec:ladder-ham}. In the
following we assume that it is only $K_{\rho}$ responsible for
$\alpha \neq 0$, while $K_{\sigma}=1$. Several different
techniques have been used to measure the $\alpha$ exponent. These
include PES and ARPES (4-150meV; 5-300K), STM(0-50meV; 10-50K),
resistivity(30-300K); where in parenthesis we have indicated
temperature and frequency ranges where the fits were performed. We
see that at least one of them is always larger than $\Lambda_{2}$
which makes our theory applicable. All probes are consistent and
gave the spectral function exponent $\alpha \approx 0.55 \div 0.6$
which means $\bar{K}_{\rho}\approx 0.24\div 0.25$. This is
reasonably close to $K_{\rho S,A}[\Lambda_2]\in (0.25, 0.29)$
predicted theoretically in Sec.\ref{ssec:4kF}, but also to
$K_{\rho}^{c}=1/4$. This shows the importance of the proximity to
the Mott transition: it can cause a strong renormalization of
$K_{\rho}$ TLL parameters towards lower values and thus observed
large value of $\alpha$ exponent.

One can be even more explicit: the only way to obtain values
$\alpha>1/2$ (which translates into $K_{\rho}<0.27$) in our model
is by assuming that the combined $4k_F$ perturbations (intra-chain
$g_3$ together with $g_{\perp}^{\pi}$ and $g_{\perp}^{u}$ terms)
can drive the RG flow and lead to strong renormalization of the
charge TLL parameters. There is one more argument which supports
this scenario. It is lack of spinon peak which has not been
observed in any PES (or ARPES) experiment in the last two decades.
This can be justified theoretically only if $\alpha>1/2$, which
provides strong limitations for the expected $\bar{K}_{\rho}^*$
values.

There is one experimental result which requires a comment. It is
the temperature dependence of $\alpha$ invoked in
Ref.\cite{JWAllen-alphaRG}, based on ARPES results. There the
following value $\alpha=0.9$ was measured at high temperatures.
This means $(K_{\rho S}+K_{\rho A})/2+(K_{\rho S}^{-1}+K_{\rho
A}^{-1})/2 \approx 6$ which translates into $\bar{K}_{\rho}^*
\approx 0.17$. This value is much smaller than our predictions. We
would like to emphasize that within TLL theory such behavior is
very unlikely. To be precise: as explained above it is not allowed
to say that these different values comes out different points
along RG trajectory. One model can possibly have only one ground
state described by $\bar{K}_{\rho}^*$. In our opinion the observed
temperature dependence can originate from a significant influence
which phonons occupancy have on the LDA band
broadening\cite{Thomas} The fact that $\alpha(T)$ dependence seems
to change with sample preparation (see Fig. 3d in
\cite{JWAllen-alphaRG}) strongly supports this interpretation.
Then the real value of $\alpha$ is the low energy one.

The other result is the estimate of a gap which according to our
reasoning may be potentially opened by the $4k_{F}$ instabilities.
However the value has been found to be extremely small $\sim
1meV$, thus our procedure, which relies on the RG analysis
starting from 200meV, is insufficient to make any definite claims.
What is more this value is much smaller than $\Lambda_2$, while
there are several effects (like disorder or the interchain hopping
$t_{\perp}$) which are potentially of order $\Lambda_2$ and
destructive for such a gap. If the hierarchy of amplitudes was
opposite (gap much larger than $t_{\perp}$) then the gap would
suppress single particle hopping. In our case the outcome is
unclear. Certainly a more sophisticated theory is necessary to
understand the unusual physics taking place below 20meV.
Experimentally this is also a controversial issue. Such a gap (in
the charge sector or in $A_{TLL}(\omega)$) has never been seen in
any experiment probing $\omega>3meV$. So far the only exception
are low temperature resistivity measurements\cite{Hussey-rhoB},
but even there a clear Arhenius-type activation behavior (at
around 1meV) is
preceded by an unusual power law like behavior. %One can only state
%that Mott gap should not affect the physics taking place above
%this energy scale.

\subsection{The validity of 1D approximation}\label{ssec:disc-tperp}

It is known\cite{giamarchi_book_1d} that the strong correlation
effects (the formation of Luttinger liquid) are able to strongly
reduce the value of $t_{\perp}$. To be more precise they strongly
reduce the energy scale where system gains coherence along c-axis.
With the value of the $\alpha$ exponent discussed above the
perpendicular hopping $t_{\perp}$, will get strongly renormalized
down to an effective value:
\begin{equation}\label{eq:tperpRG}
    t_{\perp}^{eff}= t \left( \frac{t_{\perp}}{t} \right)^{\left( \frac{1}{2-\zeta} \right)}
\end{equation}
where $\zeta=\alpha+1$ is a single particle Greens function
exponent, and $\eta_{t\perp}=2-\zeta$ is a scaling dimension of
the hopping operator. This gives a suppression of $t_{\perp}$ by a
factor $\approx 20$. To be precise this means that due to the
presence of strong interactions the hopping in the perpendicular
direction becomes coherent only $\approx 1meV$.

On the top of it there is also another source of $t_{\perp}$
renormalization, which originates from the competition with the
terms $V_{out}(q=2k_{F})$ (as described in Sec.\ref{ssec:2kF}).
The $2k_F$ terms which are a functional of a charge asymmetric
mode $\phi_{\rho A}$(e.g. the one inducing a $\pi-$DW) tend to
suppress $t_{\perp}$ \cite{Suzumura-t-perpRG}. Their influence is
non-zero only when spin sector becomes coherent and $t_{\perp}$ is
sufficiently small (there is a Bessel function
$J_{0}(t_{\perp}[l])$ involved, which arises in a very similar way
like the one in Eq.\ref{eq:RG-Krho-umkl}). It was shown
\cite{Suzumura-t-perpRG2, Suzumura-t-perpRG} that the significant
suppression of $t_{\perp}$ may happen only when
$t_{\perp}[l]\approx V_{out}(q=2k_{F})$, which (according to
scaling given in Eq.\ref{eq:tperpRG}) can be the case in our
problem but only for energies below $\Lambda_2$.

The frustration of the nearest and the next-nearest perpendicular
hopping also suppress the coherence of perpendicular hopping also
on second order when e.g. particle-hole processes are considered.
This is immediately visible if one writes a formula for any
susceptibility $\chi(q,q_{\perp},\omega)$ within the mean field,
RPA level:

\begin{equation}\label{eq:suscep-perp}
    \chi(q,q_{\perp},\omega)=\frac{\chi^{TLL}(q,\omega)}{1+(2t_{\perp}\cos(q_{\perp}c/2)+2t'_{\perp}\cos(q_{\perp}c))\chi^{TLL}(q,\omega)}
\end{equation}

where we are using simplified version of perpendicular dispersion
$\varepsilon(\vec{q_{\perp}})$ known from
Eq.\ref{eq:tight-bind-disp-def}. We see that the sign difference
between $t_{\perp}$ and $t'_{\perp}$ can cause a suppression of a
second term in the denominator. This obviously weakens the
$q_{\perp}$ dependence of susceptibility (thus also a
corresponding observable), but also has its influence if one
wished to develop a perturbative series to study the influence of
$t_{\perp}$.

Finally there are also disorder effects which can localize
carriers in the perpendicular direction. They are described in the
next section.

From experiments (optical spectroscopy, Kadowaki-Woods ratio) we
know that 1D physics seems to be correct even down to 2meV. In the
light of the above discussion these statements are not
unreasonable. Thus, the assumption about validity of 1D physics
should be valid even down to few meV's and certainly in the "high
energy" range of energies (200-20meV) this paper is dedicated for.
The values of TLL parameters indeed strongly support the idea that
the 1D regime should be able to persist to temperatures much lower
than the bare perpendicular hopping.

\section{How pertinent is substitutional disorder?}\label{ssec:disc-disor}

The most likely source of a disorder in purple bronze are random
vacancies of $Li$ atoms. From the DFT results\cite{Thomas} we know
that energy shifts caused in this way within dispersive bands can
be at most 15meV (when the $Li$ are completely removed). This sets
the strength of the substitutional disorder potential. Because
$Li$ atoms are placed well outside zig-zag chains it is reasonable
to assume the Coulomb potential interaction ($\sim q^2$) between
impurity and TLL, with predominant small q-exchange scattering.
Thus the disorder will have primarily forward character and $D_f
\approx 15meV$.

Thus we can assume the model of TLL with a forward disorder to
check if it explains the observed temperature anomalies of ARPES
below 200K. The spectral function for forward disorder is known in
real space:
\begin{equation}\label{eq:A-disord}
    A(x,t)=A_{TLL}(x,t)\exp\left(-\frac{D_{f}K_{\rho}^2}{u_{\rho}^2}x \right)
\end{equation}
where $A_{TLL}$ is the pure Luttinger liquid spectral function,
which is known \emph{e.g.} see Eq.\ref{eq:LL-propag}. The behavior
of the Fourier transform is easy to extract in the limits of large
and small $\omega$. For the high-energy (15-150meV) range we
expect the $A_{TLL}$ power law scaling (because we work above the
energy range disorder can affect). This is definitely not seen in
experimental data \cite{JWAllen-Tscal}. For the low energies
(2-15meV) we expect a convolution of standard TLL signal with an
exponential decay. A broadening of $A_{TLL}$ was indeed observed,
but it was suggested\cite{Allen-latest} that the Gaussian
convoluted with the LL $A_{TLL}(q,\omega)$ provides better fit of
the ARPES data. If one looks back how the eq.\ref{eq:A-disord} was
derived she/he finds that the approximation of uncorrelated
scattering events is not obeyed. The presence of a Gaussian is
like if the scattering events were not random, but momentum
conserving. There is also discrepancy on the level of energy
scales, experimentally the broadening has been observed already at
20meV which is larger than the maximal amplitude of disorder
15meV.

Let us now move to discussion of the lowest temperatures effects.
By analogy with a reasoning in Sec.\ref{ssec:inter-values} we
expect that $D_f$ is accompanied with a component which provides
source of random scattering events with large momentum exchange.
Its amplitude should be one order of magnitude smaller and thus
could potentially affect only the lowest energy scales. Taking
into account the $\alpha$ exponent value (and repulsive character
of interactions $K_{\rho}\ll 1$) the backward disorder, if
present, has to be highly relevant. It is known that in 1D systems
the backward disorder, in the form of cosine operator, when
present kills the superconductivity (of the BCS type).
Experimentally the unusual ARPES was observed precisely for the
superconducting samples. As expected for more disordered samples
the superconductivity disappeared, while the broadening of ARPES
signal around 100K was unchanged. The disorder also does not fit
well neither with magnetoresistivity nor STM experiments. It is
also claimed (from optical spectroscopy) that the mass of
remaining mobile carriers decreases below 30K.

In conclusion this simplest notion of disorder is incompatible
with observed effects. However the discussion is not yet closed.
The strength of the disorder potential $D_f$ is comparable with
the zero point motion or the thermal expansion effects, which will
certainly affect its influence. The crystal structure is quite
complex and supports much more sophisticated mechanisms
\emph{e.g.} the relative rotations of different $Mo$ octahedra.
These rotations provide another source of disorder in the
material. However from the overall analysis of the structure it is
certain that all these microscopic effects should affect
$t_{\perp}$ stronger than the inside-chain Luttinger liquid
physics. What is more $t_{\perp} \approx D_f$ which means that
relative strength of disorder is quite large.

\section{Conclusion}\label{sec:concl}

We studied the low energy physics of a quasi-1D material, lithium
molybdenum purple bronze. Already before it has been shown that
this material is extremely anisotropic, there is a linear
dispersion along b-axis extending down to  0.4eV, while in the
perpendicular direction dispersion is at least two orders of
magnitude weaker. In this study the physics is expressed in terms
of field theory. Our work cover the energy range where the physics
of charge modes, holons describes well the dynamics of the
compound. It begins at around 0.2eV where the experiment has
clearly shown an emergence of 1D spectral properties, in
particular the fermionic bands seem to merge at this energy scale
giving rise to a single entity. The regime of our interest extends
as low as 1D physics is strictly valid. To be on the safe side we
set it at around 20meV which is larger than any of the possible
disturbances.

The starting point of our study is the recent LDA-DFT result where
the peculiar band structure of the material has been re-confirmed.
Based on this we construct the effective low energy theory: the
band structure around Fermi energy is casted into a tight-binding
model. In addition a minimal model has to contain the strong
correlation terms, laying beyond LDA-DFT approximation. The aim of
the next few section is to provide the quantitative description of
these strong interactions.

We begin with parameterizing the interactions taking place inside
a single zig-zag chain. From previous experimental and numerical
works we are able to extract an effective real space model with a
physically reasonable values of strong correlation parameters. Due
to sparse arrangement of the chains the interactions have a finite
range character, however in the first approximation we use U-V
model to obtain the values of Luttinger liquid parameter
$K_{\rho}$. This estimate is based on several numerical works
dedicated for models which are very similar to ours. All values
converge at $K_{\rho}\approx 1/3$, which is rather low value, in
particular it may allow $4k_F$ instabilities to dominate the
physics. It also suggests that the umklapp processes are at play.
In addition we also investigate the effects which can arise if one
goes beyond simplistic U-V model. A separate section is dedicated
to a spin sector, where we estimate basic energy scales.

Later parts of this work are devoted to inter-chain physics. As
the chains are grouped in a well distinguishable pairs, it is
tempting to propose a description within a ladder-like model. We
use dielectric approximation in order to estimate strength of
inter-chain interactions, considering processes with both small
and large momenta exchange. With this knowledge we propose that
the description of physics in the considered charge regime should
be done within framework of Luttinger liquid consisting of four
modes, the two charge modes corresponds to the total (symmetric)
and the transverse (asymmetric) fluctuations. This description
fully incorporates the inter-chain processes with small momentum
exchange together with the intra-chain physics.

The processes with large momentum exchange can be taken as
perturbation and treated within RG approach. However, we claim
that in order to achieve a valid description of the system one has
to consider the intra-chain umklapp together with them. We derive
a full system of RG equations which cover intra- as well as
inter-chain instabilities and study possible trajectory of the
flow. This allows us to develop a description of purple bronze
down to the limits of validity of 1D theory. In the discussion we
show that these limits can be safely extended to energy scales two
or even three times smaller than 20meV. This allowed us to make a
more extensive, quantitative comparison with experimental results,
which basically confirms our theoretical insight into this
complicated compound.

We have achieved the effective low energy description of
$Li_{0.9}Mo_{6}O_{17}$ compound which is able to explain
convincingly experimental findings down to 20meV. In particular we
showed that for the specific combination of parameters, which are
present in this material, an unusual situation may occur. Due to
their mutual competition, the critical phase (the Luttinger
liquid) is able to survive down to the very low energy scales.
Despite the two leg ladder formation, no gap opens in holons
spectra (at least not above 20meV). This is contrary to the usual
case where a significant gap is present in a transverse
(asymmetric) mode of a ladder-like low-dimensional system. Such an
unusual physics is in agreement with the physics extracted from
experimental investigations.

It is likely that for the lowest energies the purple bronze falls
into category of doped Mott insulators with extremely small gap,
but further developments of the theory are necessary to make any
definite claims about this highly interesting regime.

\appendix

\section{Derivation of tight-binding parameters}\label{app:tight-bind}

The central result of DFT calculation is a band structure of a
given material. As it contains a huge amount of information
usually it is difficult to deal with when the effective, low
energy theory is constructed. In such a case a standard procedure
is to approximate a solid by a tight binding model with a selected
sites located at their positions $r_{i}$ and unknown hopping
parameters between them. Usually only the nearest $t$ and the
next-nearest $t'$ neighbor hoppings are taken into account. We are
interested in the low energy physics, thus the aim is to fit bands
crossing Fermi energy $E_F$ (in the given direction $d$) with the
properly chosen parameters $t_d$ $t'_d$.

As described in Sec.\ref{sec:bands} in the case of $Li_{0.9}Mo_6
O_{17}$ most of low energy spectral weight in located on $Mo(1)$
and $Mo(4)$ sites (see Fig.\ref{fig:structure}a)) so we take only
them into further considerations. As explained only two bands
cross $E_F$ thus our aim is to fit these two dispersions.

The dispersion relation for tight-binding model defined above is
known:
\begin{multline}\label{eq:tight-bind-disp-def}
    \varepsilon(\vec{k})=-t\cos(k_{b}b/2)-t'\cos(k_{b}b)\\
    - \frac{t_{\perp 1} + t_{\perp 2}}{2}\cos(k_{c}c/2)-\frac{t_{\perp 1} - t_{\perp
    2}}{2}\sin(k_{c}c/2)\\
    - \frac{t'_{\perp 1} + t'_{\perp 2}}{2}\cos(k_{c}c)-\frac{t'_{\perp 1} - t'_{\perp
    2}}{2}\sin(k_{c}c)
\end{multline}
where the first line describes the dispersion along b-axis, the
last two the dispersion along c-axis (with the dimerization
included). We neglected dispersion along a-axis. For the b-axis
dispersion we assumed the simplest tight-binding model, as the LDA
result suggest rather straightforward interpretation of the bands.
We introduced next nearest neighbor hopping $t'$ mostly because
the zig-zag chain structure seems to allow for this refinement,
however from the general shape of bands in $\vec{b}$ direction it
does not seem to be necessary.

For the c-axis dispersion the situation is quite different. We
introduced much more terms because the curve is quite unusual,
with a well pronounced double minima and a node at $k_c =0 $, a
feature quite difficult to fit within standard model. Based on the
analysis of the structure presented in Fig.\ref{fig:structure}a)
we deduce that:
\begin{description}
    \item[i] there are two times more intra-ladder than
    inter-ladder links (they are linked either directly through $Mo(1)-Mo(4)$ bond or auxiliaries through $Mo(2)-Mo(5)$ bond)
    \item[ii] the inter-ladder hopping goes always though $Mo(2)$ (or $Mo(5)$)
    atoms, these are two paths which can interfere. These hoppings
    are possible only every second site.
    \item[iii] the next-nearest neighbor hopping is allowed only
    through $Mo(2)-Mo(5)$ thus should be much smaller than the one
    above. These hoppings are also possible only every second
    site.
\end{description}
all of the hoppings goes either via a $\delta$-bonds or
$\pi$-bonds (the second are allowed only due to octahedra
tilting). None of the links should be very much stronger than the
others. From the crystal structure analysis all other hopping
should be negligible, thus the task is to fit the dispersion with
the above given model Eq.\ref{eq:tight-bind-disp-def}.

\begin{figure}
  \includegraphics[width=\columnwidth]{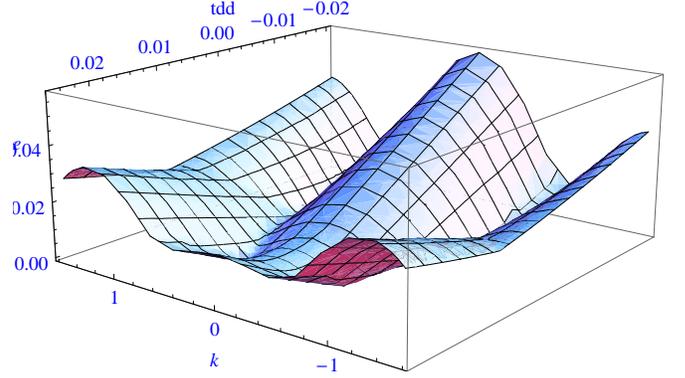}\\
  \caption{The dispersion along c-axis shown for varying direct $\delta$ hopping along $Mo(1)-Mo(4)$ bond.
  The strength of the other hopping path (through $Mo(2)-Mo(5)$) is taken as 0.02 while the inter-ladder one 0.01.}\label{fig:dispers}
\end{figure}

On Fig.\ref{fig:dispers} we show an example of $k_c$ dispersion
with conditions [i]-[iii] fulfilled. We see that it is possible to
obtain quite similar shape to the LDA one provided that two
interfering paths have the hopping parameters of the opposite
sign. The only quantitative issue would be relatively large value
of band splitting found in Ref.\cite{Thomas}. It may appear from
Hartree interaction term if the two bands had different orbital
character. One should also remember that the value of splitting
found in \cite{Popovic-bron-DFT} was a bit smaller. We will leave
this issue for further specialized studies like \emph{e.g.} NMTO.

These findings are summarized in a table Tab.\ref{tab:tight-bind}

\begin{table}
  \centering
  \caption{The values of parameters (given in meV) of the tight binding hamiltonian
  given in Eq.\ref{eq:tight-bind-disp-def} which fits best
  the LDA result\cite{Thomas}, the dispersion along c-axis.}\label{tab:tight-bind}
  \begin{tabular}{|c|c|c|c|}
    \hline
    $t_{\perp 1}$ & $t_{\perp 2}$ & $t'_{\perp 1}$ & $t'_{\perp 2}$ \\
    \hline
     15 & 5 & -10 & $<5$ \\
    \hline
  \end{tabular}
\end{table}

\section{Derivations of cosine terms}\label{app:cos-deriv}

We begin with a fermionic for of considered scattering processes:
1. the intra-chain umklapp:
\begin{multline}\label{eq:umklapp-ferm}
    H_{cos3}^i = g_3 \sum_{q_{i}}(\psi_{+q_{1}i\uparrow}^{\dag}\psi_{+q_{2}i\downarrow}^{\dag}\psi_{-q_{3}i\downarrow}\psi_{-q_{4}i\uparrow}+\\
    \psi_{-q_{1}i\uparrow}^{\dag}\psi_{-q_{2}i\downarrow}^{\dag}\psi_{+q_{3}i\downarrow}\psi_{+q_{4}i\uparrow})
\end{multline}

2. Interchain umklapp scattering ($4k_F$):
\begin{multline}\label{eq:umklapp-boson}
    H_{cos3}^{\perp} = g_3^{\perp} \sum_{q_{i}}(\psi_{+q_{1}1\uparrow}^{\dag}\psi_{+q_{2}2\downarrow}^{\dag}\psi_{-q_{3}2\downarrow}\psi_{-q_{4}1\uparrow}+\\
    \psi_{-q_{1}1\uparrow}^{\dag}\psi_{-q_{2}2\downarrow}^{\dag}\psi_{+q_{3}2\downarrow}\psi_{+q_{4}1\uparrow})
\end{multline}

where the following relation between momenta holds
$q_{1}+q_{1}-q_{1}-q_{1}=0$ in order to preserve momentum
conservation during scattering; the low energy limit implies that
for each momenta there exist the ultraviolet cut-off
$|q_{i}|<\Lambda$.

3. Interchain exchange scattering ($4k_F$):
\begin{equation}\label{eq:umklapp-boson}
    H_{cos}^{\pi} = g_{\pi}^{4k_{F}} \int_{0}^{L}dx \rho_1^{4k_F}\rho_2^{4k_F}
\end{equation}

4. Interchain exchange scattering ($2k_F$):
\begin{equation}\label{eq:umklapp-boson}
    H_{cos}^{\pi} = g_{\pi}^{2k_{F}} \int_{0}^{L}dx \rho_1^{2k_F}\rho_2^{2k_F}
\end{equation}

where $\rho_i^{2k_F}$ is the $2k_F$ charge density in the
\emph{i-th} chain.

Usually one is also interested in correlation functions, which
have the general form:
\begin{equation}\label{eq:corr_gener}
    R_{O}=<T_{\tau}O(x,t)O^{\dag}(0,0)>
\end{equation}
where the simplest intra-chain examples of possible operators $O$
are:
\begin{itemize}
    \item in Peierls channel - the charge density wave (CDW)\\
    $O_{CDW}(x,t)=\sum_{r,\sigma,\sigma'}\psi_{r}(x,t)^{\dag}\delta_{\sigma\sigma'}\psi{\bar{r}}(x,t)$
    \item in Cooper channel - the singles superconductivity (SS)\\
    $O_{SS}(x,t)=\sum_{r,\sigma,\sigma'}\sigma\psi_{r}(x,t)\delta_{\sigma\bar{\sigma'}}\psi_{\bar{r}}(x,t)$
\end{itemize}

When we consider a single chain then a chiral fermion creation
operator is related to bosonic spin and charge fields as follows
(in the continuum limit):
\begin{widetext}
\begin{equation}\label{eq:ferm-to-boson}
    \psi_{r,\sigma}(x)=\frac{1}{\sqrt{2\pi\alpha}}\eta_{r,\sigma}\exp(\imath
    rk_{F}x)\exp-\imath/\sqrt{2}(r\phi_{\rho}(x)-\theta_{\rho}(x)+\sigma(r\phi_{\sigma}(x)-\theta_{\sigma}(x)))
\end{equation}
where the coefficient $\eta_{r,\sigma}$ is the Majorana fermion
which doesn't have any spatial dependance and it is introduced
only in order to preserve anti-commutation for the fermion
operators $\psi$. Usually they don't play any role in the physical
description of the system but they are able to change signs of
some correlation functions, so one has to take care about them.

In the case of the ladder one has a straightforward
generalization:

\begin{equation}\label{eq:ferm-to-bos-ladder}
    \psi_{r,\sigma,\nu}(x,t)\sim\eta_{\sigma ,\nu}\exp(\imath r k_{F}x)\exp[-\frac{\imath}{2}(r\phi_{+\rho}+\theta_{+\rho}+\sigma (r\phi_{+\sigma}+\theta_{+\sigma})+\nu (r\phi_{-\rho}+\theta_{-\rho}+\sigma (r\phi_{-\sigma}+\theta_{-\sigma})))]
\end{equation}

\end{widetext}

The above given equations Eq.\ref{eq:ferm-to-boson} and
Eq.\ref{eq:ferm-to-bos-ladder} allow to rewrite all fermionic
terms in the hamiltonian like \emph{e.g.} the one in
Eq.\ref{eq:umklapp-ferm} or in general any interesting operator
into the language of bosonic fields.  In our particular case we
are interested in the following interaction terms:

1. Umklapp scattering (at quarter filling) in the \emph{i-th}
chain
\begin{equation}\label{eq:umklapp-boson}
    H_{cos3} = g_3 \int_{0}^{L}dx \cos(2\sqrt{8}\phi_{i \rho}(x)+\delta)
\end{equation}
where $\delta$ indicates the doping away from a commensurate case,
quarter filling in our case (where $\exp\imath(\pi- 4 k_{F})x$ is
not oscillating).

2. Interchain umklapp scattering ($4k_F$):
\begin{equation}\label{eq:umklapp-boson}
    H_{cos3}^{\perp} = g_3^{\perp} \int_{0}^{L}dx \cos(4\phi_{S \rho}(x)+\delta)
\end{equation}

3. Interchain exchange scattering ($4k_F$):
\begin{equation}\label{eq:umklapp-boson}
    H_{cos}^{\pi} = g_{\pi}^{4k_{F}} \int_{0}^{L}dx \cos(4\phi_{A \rho}(x))
\end{equation}

4. Interchain exchange scattering ($2k_F$):
\begin{equation}\label{eq:umklapp-boson}
    H_{cos}^{\pi} = g_{\pi}^{2k_{F}} \int_{0}^{L}dx \cos(2\phi_{A \rho}(x))\cos(\sqrt{2}\phi_{1 \sigma}(x))\cos(\sqrt{2}\phi_{2 \sigma}(x))
\end{equation}

These are the terms for which RG equations are derived in
Sec.\ref{sec:RG}. In all the above we took a following convention:
\begin{equation}\label{eq:fields-relations}
    \phi_{S,A}(x)=\frac{\phi_1(x) \pm \phi_2(x)}{\sqrt{2}}
\end{equation}
which allows to go from one basis to another.

\section{Initial value of umklapp terms}\label{app:bare-umkl}

Estimating the value of $g_3$ at certain energy scale $\Lambda_0$
is an impossible task, so all our results cannot be taken strictly
quantitatively. This is a generic problem of the RG method present
for any model even a simple half filled chain. The complexity of
our system, makes the task even more tedious.

We propose the approach, based on the first order expansion of RG
equations, to get a reasonable value of $g_{3}(\Lambda_0)$.
However one has to keep in mind that as we are are working with
rather large couplings, of order 0.1 (but always smaller than
0.25) so higher order terms can introduce non-negligible
corrections even within single instability flow. Thus the results
has to be taken with caution, can be thought as only
approximation.

With this remarks being said we can proceed. One can in principle
integrate out BKT equations to get a values of along the flow. One
gets the following result\cite{giamarchi_book_1d}:
\begin{equation}\label{eq:RG-itegr1}
    g[l]= \frac{A}{\sinh\left[ Al+\tanh^{-1}(A/y_{\parallel}^0)\right]}
\end{equation}
\begin{equation}\label{eq:RG-itegr1}
    y_{\parallel}[l]= \frac{A}{\tanh\left[ Al+\tanh^{-1}(A/y_{\parallel}^0)\right]}
\end{equation}
where A is an invariant of the flow, in our case $A=1/12$. Every
point on an RG trajectory leads us to fixed point value
$K_{rho}^*$. From numerics based on U-V model we found
(Sec.\ref{sec:LL-res}) $K_{rho}^*=1/3$ and we know that for
quarter-filled chain $K_{rho}^c=1/4$. Thus in our case
$A=K_{rho}^*-K_{rho}^c=1/12$. Our aim is to extract the value of
$K_{rho}[l]=K_{rho}^c+y_{\parallel}[l]$ at certain energy scale
$\Lambda_1$ (corresponding to $\approx 0.2 eV$) which upon RG
leads to known $K_{rho}^*$. This procedure can be thought as
moving against the direction of the RG flow.

The only missing quantity in Eq.\ref{eq:RG-itegr1} is
$y_{\parallel}[l=0]$ which should be interpreted as the distance
of an initial point of the flow $K_{rho}[l=0]$ (formally in the
infinite energy, i.e. somewhere around UV-cut-off of the model)
from $K_{rho}^{c}=1/4$. Although around the fixed point the flow
is rather vertical, we decided to take the most conservative (and
giving the most modest value of $g_{3}(\Lambda_0)$) assumption
that $K_{rho}[l=0]=1/2 \Rightarrow y_{\parallel}[l=0]=1/4$. It is
because certainly at $K_{rho}=1/2$ the quarter filled chain
remains insensitive to Mott localization.

With this we are able to estimate the bare
$g_{3}(\Lambda_1)\bar{\Lambda}$ to be around $0.35eV$. Then
$y_{\parallel}(\Lambda_1)$ is only 6\% larger than corresponding
$g(\Lambda_1)$, while the position of the separatrix of RG flow in
the first order is set by $y_{\parallel}=g$. The distance from the
separatrix is of the same order like additional intra-chain terms
described in Sec.\ref{ssec:intra-rest}. From the remarks above we
know this is only an estimate, but it certainly implies that the
umklapp, even in intra-chain U-V approximation, is an order of
magnitude larger than any of $V_{\perp}(q=4k_{F})$. Although it
seems to be irrelevant, but its bare amplitude is large and quite
close to separatrix so the intra-chain umklapp has to be taken
with care in the RG analysis below.

The fact that bare $g_3^0$ falls very close to separatrix can be
confirmed by an independent reasoning. The position of separatrix
is frequently constrained by symmetry considerations. Taking into
account the hierarchy of energy scales, we take simplistic
approximation of spinless fermions in $U\rightarrow \infty$ (like
in Sec.\ref{sec:LL-res}). This obviously overestimates $g_3$, but
numerics \cite{Mila-Zotos-num} shows us that beyond a threshold
$U=4t$, the dependence on U is weak, thus the corrections breaking
(particle-hole) symmetry should be quite small. The advantage is
that now one can map charge sector on pseudo-spin model (empty and
occupied sites), for which the position of separatrix of RG flow
(phase transition) in terms of bare parameters is known exactly
thanks to SU(2) symmetry. It is located at $V_{in}^{eff^c}=2
t^{eff}$ where $V_{in}^{eff}$ accounts for all unscreened
interactions along the chain (while $t^{eff}\approx t$, if one
want to be more precise he would found it slightly reduced with
respect to $t$ because of dimerization). If we put values found
before (in table Tab.\ref{tab:str-coupl-m} for $V_{in}^{eff}$) we
indeed find ourself very close to $V_{in}^{eff^c}$.

%\bibliographystyle{plain}
%\bibliography{bronze-LL}

\end{document}